\documentclass[11pt,journal,onecolumn]{IEEEtran}

\usepackage{amsmath,etoolbox}
\usepackage{amsfonts}
\usepackage{amssymb}
\usepackage{theorem}
\usepackage{graphicx}
\usepackage{psfrag}
\usepackage{enumerate}

\usepackage{graphicx}
\usepackage{caption}
\usepackage{subcaption}
\usepackage{color}
\usepackage{bbm}

\usepackage{cite}

\usepackage[mathscr]{euscript}

 \usepackage{makecell}

 \usepackage{setspace}
\newtheorem{theorem}{Theorem}
\newtheorem{lemma}{Lemma}

\newtheorem{corollary}{Corollary}
\newtheorem{definition}{Definition}
\newtheorem{remark}{Remark}
\newtheorem{example}{Example}
\newtheorem{claim}{Claim}

\newcommand{\comment}[1]{}

\newcommand{\Jithin}[1]{{\leavevmode\color{red}[#1]}}

\newcommand{\revd}[1]{ \textcolor{black}{ #1} }

\newcommand{\rev}[1]{ \textcolor{black}{ #1} }

\newcommand{\revn}[1]{ \textcolor{black}{ #1} }

\newcommand{\RNzerre}{\textcolor{black}{\eR^{RN}_{(0)}(f,X,Y)}}

\newcommand{\RNzerun}{\textcolor{black}{\cR^{RN}_{(u)}(f,X,Y)}}

\newcommand{\RNzerunA}{\textcolor{black}{\eR^{RN}_{(0)}(f_1,X,Y)}}
\newcommand{\RNzerunB}{\textcolor{black}{\eR^{RN}_{(0)}(f_2,X,Y)}}

\newcommand{\RNeps}{\textcolor{black}{\eR^{RN}_{(\epsilon)}(f,X,Y)}}
\newcommand{\RNepsA}{\textcolor{black}{\eR^{RN}_{(\epsilon)}(f_1,X,Y)}}
\newcommand{\RNepsB}{\textcolor{black}{\eR^{RN}_{(\epsilon)}(f_2,X,Y)}}
\newcommand{\RNepsX}{\textcolor{black}{\eR^{RN}_{(\epsilon)}(\CWOO,X,Y)}}

\newcommand{\BFNzer}{\textcolor{black}{R^{*(BFN)}_{(0)}(f,X,Y)}}

\newcommand{\BFNeps}{\textcolor{black}{R^{*(BFN)}_{(\epsilon)}(f,X,Y)}}

\newcommand{\frooks}{\textcolor{black}{RG^{f}_{XY}}}
\newcommand{\frookA}{\textcolor{black}{RG^{f_1}_{XY}}}
\newcommand{\frookB}{\textcolor{black}{RG^{f_2}_{XY}}}

\newcommand{\frooksun}{\textcolor{black}{RG^{f,(u)}_{XY}}}

\newcommand{\faux}{\textcolor{black}{\widetilde{RG}^{f}_{U_1U_2}}}

\newcommand{\CWOO}{\textcolor{black}{CWOOF} }

\def\cX{\mbox{$\cal{X}$}}
\def\cY{\mbox{$\cal{Y}$}}
\def\cZ{\mbox{$\cal{Z}$}}
\def\cU{\mbox{$\cal{U}$}}
\def\cV{\mbox{$\cal{V}$}}
\def\cR{\mbox{$\cal{R}$}}

\def\eR{\mbox{$\mathscr{R}$}} 

\def\cG{\mbox{$\cal{G}$}}

\def\cA{\mbox{$\cal{A}$}}

\def\cC{\mbox{$\cal{C}$}}

\def\cQ{\mbox{$\cal{Q}$}}

\title{Function Computation through a Bidirectional Relay} 
\author{
Jithin~Ravi and Bikash~Kumar~Dey
\thanks{This paper was presented in part at the IEEE Information
Theory Workshop (ITW), Jerusalem, Israel, April 2015 and at the IEEE GLOBECOM NetCod 2016, Washington, DC, USA, December 2016.
This work was supported by the Department of Science and Technology under grant SB/S3/EECE/057/2013
and by Information Technology Research Academy under grant ITRA/15(64)/Mobile/USEAADWN/01.

J.~Ravi and B.~K.~Dey are with the Department of Electrical Engineering at
 IIT Bombay, Mumbai, INDIA-400076. Email:\{rjithin,bikash\}@ee.iitb.ac.in.
}

}

\begin{document}
    \maketitle
\begin{abstract}
We consider a function computation problem in a three node wireless network. Nodes A and B observe
two correlated sources $X$ and $Y$ respectively, and want to compute a function $f(X,Y)$. 
To achieve this, nodes A and B
send messages to a relay node C at rates $R_A$ and $R_B$ respectively. 
The relay C then broadcasts a message to A and B at rate $R_C$.
We allow block coding, and study the achievable region of rate triples
under both zero-error and $\epsilon$-error.
As a preparation, we first consider a broadcast network 
from the relay to A and B. A and B have side information $X$ and $Y$ respectively.
The relay node C observes both $X$ and $Y$ and broadcasts 
an encoded message to A and B.
We want to obtain the optimal broadcast 
rate such that A and B can recover the function $f(X,Y)$
from the received message and their individual side information $X$ and $Y$ respectively.
\rev{
For this problem, we show equivalence between $\epsilon$-error and zero-error computations-- this 
gives a rate characterization for zero-error computation.
As a corollary, this also gives 
a rate characterization for  the relay network under zero-error 
for a class of functions called {\em component-wise one-to-one functions}
when the support set of $p_{XY}$ is full. 
}
For the relay network, the zero-error rate region for arbitrary functions is characterized in terms of
graph coloring of some suitably defined probabilistic graphs. We then
give a single-letter inner bound to this rate region. Further, we extend the graph theoretic
ideas to address the $\epsilon$-error problem and obtain a single-letter inner bound.
\end{abstract}


\section{Introduction}
Distributed computation of distributed data over a network has been investigated
in various flavours for a long time. Gathering all the data at the
nodes where a function needs to be computed is wasteful in most situations.
So intermediate nodes also help by doing some processing of the data
to reduce the communication load on the links. Such computation frameworks
are known as distributed function computation or in-network function
computation \cite{Korner_1979,Han_1987,Orlitsky_2001,Kowshik_2012,Rai_2012,Shah_2013}.

\begin{figure}[htbp]
\centering
\includegraphics[scale=0.5]{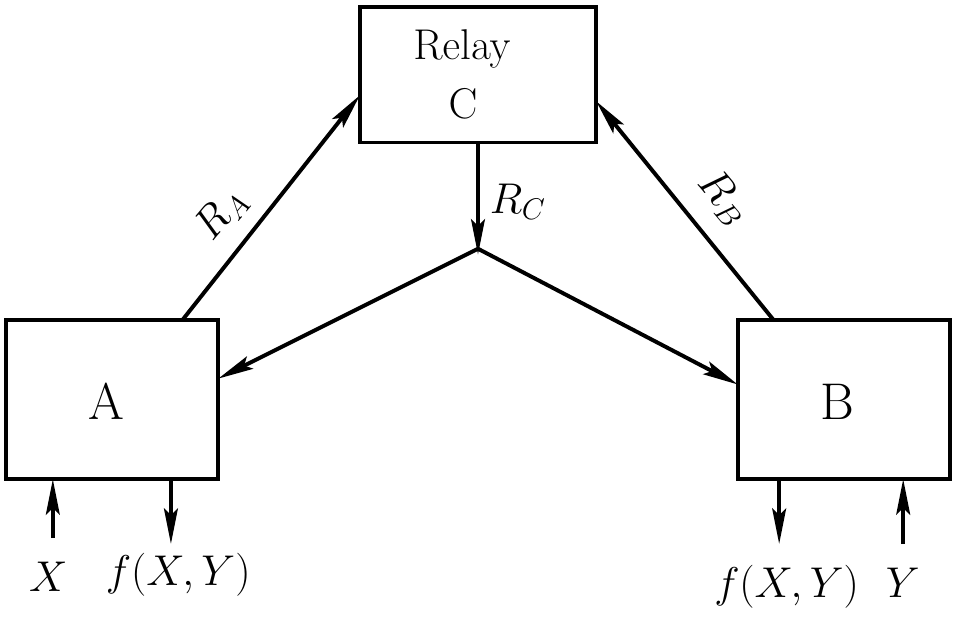}
\caption{Function computation in bidirectional relay network (RN)}
\label{Relay_Network}
\end{figure}

We consider the problem of function computation in a wireless relay network (RN) with
three nodes as shown in Fig.~\ref{Relay_Network}.
Nodes A and B have two correlated random variables $X$ and $Y$ respectively.
They have infinite i.i.d. realizations of these random variables.
They can communicate directly to a relay node C over
\revd{independent} error-free links. The relay node C can send a message to
both A and B over a noise-less broadcast link. 
Nodes A and B want to compute a function \revd{$f(X,Y) = Z$}.
We allow block coding of arbitrarily
large block length $n$. We allow
two phases of communication. In the first phase, both A and B
send individual messages to C at rates $R_A$ and $R_B$ over the 
respective \revd{independent} links.
In the second phase, the relay broadcasts a message to A and B at rate $R_C$.

The broadcasting relay in the model captures one aspect of wireless networks.
We consider our function computation problem over this network
under zero-error and $\epsilon$-error criteria.
Under zero-error, both nodes want to compute the function with no error.
Under $\epsilon$-error, the probability of error in computing the function 
should go  to zero as block length tends to infinity.
A special case of this problem have been studied in~\cite{Wyner_2002,Su_2010}.
Exchanging $X$ and $Y$
was considered in ~\cite{Wyner_2002}, and the rate region was characterized
in the $\epsilon$-error setting. For this problem, some single-letter 
inner and outer bounds were given for the rate-distortion function
in~\cite{Su_2010}.

\begin{figure}[h]
\centering
\includegraphics[scale=0.5]{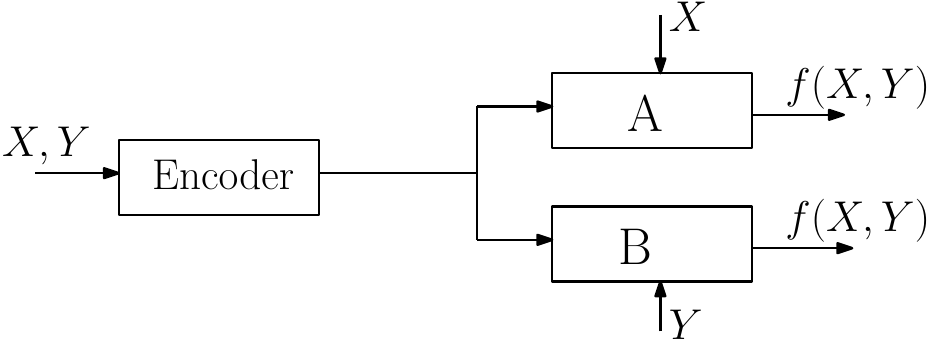}
\caption{Broadcast function network with complementary side information (BFN-CSI)}
\label{Broadcast_network}
\end{figure}

As a preparation to address the problem in Fig.~\ref{Relay_Network},
we first consider {\em the broadcast function network with complementary
side information} (BFN-CSI) shown in Fig.~\ref{Broadcast_network}.
This problem arises as a special case of the function computation problem in the relay network,
when A and B communicate $X$ and $Y$ to the relay node. 
In the relay network, rate $R_C$ attains its minimum when the 
relay has $X$ and $Y$.
So the optimal broadcast rate for the problem in Fig.~\ref{Broadcast_network}
is the minimum possible rate $R_C$ in the relay network.
For the broadcast function network, the optimal $\epsilon$-error rate 
can be shown to be $\max \{H(Z|Y), H(Z|Y)\}$
using the Slepian-Wolf result.
We study this problem under zero-error criteria.

The problem of zero-error source coding with receiver side information 
was first studied for fixed length coding by Witsenhausen in 
\cite{Witsen_1976} using a ``confusability graph'' $G_{X|Y}$. 
The minimum rate was characterized in terms of
the chromatic number of its AND product graphs $G_{X|Y}^{\wedge n}$.
The same problem was later considered in \cite{Alon_1996}
under variable length coding, and 
the minimum rate was shown to be the limit of the normalized chromatic
entropy of $G_{X|Y}^{\wedge n}$. 
This asymptotic rate was later 
shown~\cite{Rose_2003} to be the 
complementary graph entropy~\cite{Longo_1973} of $G_{X|Y}$.
However, A single-letter characterization for complementary 
graph entropy is still unknown. 

In the absence of a single-letter characterization of  zero-error
source coding problems, many authors have studied their problems
under a stricter decoding requirement, known as the
``{\it unrestricted input}'' setup~\cite{Alon_1996,Koulgi_2003,Shayevitz_2014}. 
In this setup, even for a source vector
which has some zero-probability components (and thus the vector itself
having probability $0$), the decoder is required
to reproduce the desired symbols for the other components of the vector.
\revd{ Unrestricted input setup was introduced in \cite{Alon_1996}, and 
for the problem addressed in \cite{Alon_1996}, 
the optimum rate under unrestricted input setup was shown to be the graph entropy of the confusability graph
which has a single letter characterization. On one hand, 
under unrestricted input setup, computation problems are sometimes
tractable when the original zero-error computation problems are not. 
On the other hand, protocols for the unrestricted input setup are clearly 
also valid protocols 
for the original zero-error decoding problem. So achievable rates under
unrestricted input setup give upper bounds on the optimal zero-error rates. 
Shayevitz \cite{Shayevitz_2014} also studied the unrestricted input
version of their problem.
}
In all these models, the unrestricted input setup is represented
by the OR product of a suitable confusability graph.
In contrast, for our function computation problem in the relay network, 
the unrestricted input setup is not represented by the OR products of the
confusability graph.


For distributed coding of two sources and joint decoding,
a single-letter characterization was given 
for the unrestricted input version in~\cite{Koulgi_2003}.
Most related recent work to our present work is~\cite{Shayevitz_2014}, where a decoder having side 
information $Z$ wants to compute a function $f(X,Y,Z)$ using a message 
encoded by a relay, which in turn receives two messages encoded by
two sources $X$ and $Y$. Single-letter inner and outer bounds were
given for the unrestricted input setup.

\comment{
 For our zero-error computation problem depicted in Fig.~\ref{Relay_Network}, we 
provide a characterization of the rate region in terms of graph coloring 
of some suitably defined graphs. We provide single-letter inner and 
outer bounds for the rate region. A sufficient condition on the joint 
distribution $p_{XY}$ is identified under which, the relay will
also be able to reconstruct $f(X,Y)$ for any scheme where A and B
reconstruct it with zero error.
}

The problem of  broadcast with side information, has been studied
extensively in the literature (see \cite{Alon_2008}-\cite{Timo_2013} and references therein).
Index coding (see~\cite{Yossef_2011}-\cite{Langberg_2011}) is a special case of  broadcast with side information, and it is related to
our work.
In index coding, a server has access to $K$ binary independent 
and uniformly distributed
random variables and the receivers 
have access to different subsets of these messages. 
Each receiver wants to recover an arbitrary subset of the messages using 
its side information and the message broadcasted by the server. 
The goal is to minimize the broadcast rate of the message sent by the server.
A computable characterization of the optimum broadcast
rate for the general index coding problem is still unknown.
\rev{ For our broadcast function network (Fig.~\ref{Broadcast_network}), instead of recovering the messages, we consider the problem of computing a function of the messages. For this problem,  the optimal $\epsilon$-error rate is $\max\{ H(Z|X), H(Z|Y) \}$ (using Slepian-Wolf result), thus it is a lower bound for the optimal zero-error rate. We show that the rate $\max\{ H(Z|X), H(Z|Y) \}$  is achievable under zero-error.}

\comment{
For the broadcast function network in Fig.~\ref{Relay_Network}, 
we first consider the special case of computing 
a class of functions called
{\em component-wise one-to-one function (CWOOF)}.
This problem is equivalent to recovering $Y$ at node A and $X$ at node B,
known as the complementary delivery problem, and is
an instance of the index coding problem with two messages.
This problem has been addressed under noisy broadcast channel in \cite{Tuncel_2006,Wu_2007,Kramer_2007}
for $\epsilon$-error recovery of the messages.
Lossy version of this problem was studied in \cite{Kimura_2007, Timo_2013}. 
For the lossless case, the optimal $\epsilon$-error rate
can be shown to be
$\max\{ H(Y|X), H(X|Y) \}$ using the Slepian-Wolf result.
We show in this paper that this rate is also achievable with
zero-error.
For any index coding problem with  {\em independent and uniformly distributed messages}, 
the equivalence between zero-error and $\epsilon$-error rates has been
shown in \cite{Langberg_2011}. Our result extends this 
to correlated sources with arbitrary distribution
in the specific case of complementary delivery.
The technique followed in \cite{Langberg_2011} does not directly extend
to correlated sources.
For arbitrary function $Z=f(X,Y)$, the optimal
$\epsilon$-error rate is $\max\{ H(Z|X), H(Z|Y) \}$ (using
Slepian-Wolf result), thus it is
a lower bound for the optimal zero-error rate. We provide
a single-letter upper bound for the optimal
zero-error rate.
}

For the relay network, we study the function
computation problem under zero-error.
Suitable graphs are defined to address the problem.
We first consider computing \revd{a component-wise one-to-one function} at both the end nodes. 
\revd{Note that computing  a component-wise one-to-one function in the relay network is the equivalent to 
exchanging $X$ and $Y$ through the relay.}
Building on our results on the broadcast function network,
we give a single-letter characterization of the rate region for 
computing a \revd{component-wise one-to-one function} when the support set of $p_{XY}$ is full. 
For arbitrary functions, we study the problem under unrestricted input 
setup and provide a multiletter characterization of the rate region. 
Then we provide a single-letter inner bound 
for this region, which is also an inner bound for the zero-error problem.

Next, we consider the function computation problem in the relay network
under $\epsilon$-error. For this problem, we use the graph theoretic ideas
developed for zero-error, to get a single-letter inner bound for the rate 
region.

\subsection{Contributions and organization of the paper}

We list the contributions of this paper below.

\begin{itemize}
 \item For the zero-error function computation problem shown in 
Fig.~\ref{Broadcast_network}, in Theorem~\ref{Thm_BFN}, 
\rev{
we show that the optimal zero-error broadcast rate is same as optimal $\epsilon$-error rate which has a 
 a single-letter characterization.}\comment{
we give a single-letter characterization
 for the optimum broadcast rate for computing a component-wise one-to-one function at nodes A and B, 
and for general functions, we give upper 
 and lower bounds for the optimum rate. 
}
Using this result, we give a single-letter
 characterization of the rate region for computing a \revd{component-wise one-to-one function} in the relay network (Fig.~\ref{Relay_Network}) 
 when the support set of $p_{XY}$ is full.
 We then argue that when $X$ and $Y$ are independent, 
 exchanging $(X,Y)$ in the relay network has the same rate region under zero-error and  $\epsilon$-error.

 \item 
 We consider the zero-error function computation problem in the relay network (Fig.~\ref{Relay_Network}) 
 under the unrestricted input setup. This setup is a more constrained version of the zero-error problem.
 We  give a multiletter characterization of the rate region under this setup as well as
 for the zero-error problem (Theorem~\ref{Thm_Rate_Region}).
 The multiletter characterization is obtained using coloring of some suitably 
defined graphs.
 Our arguments based on coloring are similar to~\cite{Shayevitz_2014}.
We show that if $p_{XY}$ has full support, then the relay can
also compute the function if
 A and B can compute it with zero-error (Theorem~\ref{Relay_function}).

 \item For the unrestricted input setup, we propose two achievable schemes whose time sharing gives a single-letter inner bound 
for the corresponding rate region (Theorem~\ref{Thm_Zero_Inner1}). 
\comment{
A single-letter outer bound for the region is presented in 
Theorem~\ref{Thm_Zero_Outer1} . We show that the inner and outer bounds in Theorem~\ref{Thm_Zero_Inner1} and  Theorem~\ref{Thm_Zero_Outer1}
 meet for a class of functions (Theorem~\ref{Thm_Zero_compl_chara}).
 }

 \item The function computation problem in Fig.~\ref{Relay_Network} is 
then addressed under $\epsilon$-error.
 We extend the graph theoretic ideas used for zero-error computation to $\epsilon$-error 
 computation.
 Similar to the two achievable schemes for zero-error computation, we 
give an inner bound for the rate region using
 two achievable schemes for $\epsilon$-error computation (Theorem~\ref{Thm_Epsilon_Inner1}). 
 The cutset outer bound is given in Lemma~\ref{Lem_Asym_Outer_Bound}. 
 \item 
 For two functions $f_1, f_2$ of $(X,Y)$, we give a graph theoretic sufficient 
condition under which the rate region for computing  $f_1$
is a subset of the rate region for computing $f_2$. This condition holds 
for both zero-error and $\epsilon$-error computations (Theorem~\ref{Thm_Rooks}).
Using this result, we give a class of functions for which the rate region  
is the same as the region for exchanging $(X,Y)$.
\end{itemize}

The organization of the paper is as follows. Problem formulations for zero-error and $\epsilon$-error
are given in Section~\ref{Model_zero_error} and in Section~\ref{Model_epsl_error} respectively. Some graph theoretic definitions
are given in Section~\ref{Model_graph_definitions}. 
We provide our results for zero-error computation in Section~\ref{Sec_zero_results}.
The $\epsilon$-error results are given in Section~\ref{Sec_Epsl_results}. The proof of the results for zero-error
computation and $\epsilon$-error computation are given in Section~\ref{sec_zero_proofs} and Section~\ref{Sec_asym_bounds} respectively.
We conclude our paper in Section~\ref{Sec_conclusion}.

\begin{table}[h]
\centering
\begin{tabular}{ |p{1cm}|p{8cm}|p{6cm}|} 
 \hline
 & \thead{Zero-error} & \thead{ $\epsilon$-error}\\
 \hline
 BFN-CSI & $\bullet$ \rev{Complete characterization (Theorem~\ref{Thm_BFN})}
   & $\bullet$ Follows directly from Slepian-Wolf results \\
 \hline
 RN & $\bullet$ Multiletter characterization (Theorem~\ref{Thm_Rate_Region}) & $\bullet$  Cutset  outer bound (Lemma~\ref{Lem_Asym_Outer_Bound})\\
 & $\bullet$  Single-letter characterization for \CWOO when support set is full  (Corollary~\ref{Cor_relay_XOR})& $\bullet$  Single-letter inner bound (Theorem~\ref{Thm_Epsilon_Inner1}) \\
 & $\bullet$ Single-letter inner bound for unrestricted inputs (Theorem~\ref{Thm_Zero_Inner1}) & \\
 & $\bullet$  A sufficient condition on $p_{XY}$ under which the relay can compute the function in any zero-error scheme (Theorem~\ref{Relay_function}) 
 
 & \\
 \cline{2-3}
 & \multicolumn{2}{l|}{$\bullet$ Graph-based sufficient condition for ``rate region for $f_1 \supseteq$ rate region for $f_2$'' (Theorem~\ref{Thm_Rooks})} \\
 \hline 
\end{tabular}
 \caption{Summary of our results}
\end{table}

\section{Problem formulation and preliminaries}
\label{sec:definitions}

Nodes A and B observe $X$ and $Y$ respectively from finite alphabet 
sets $\cX$ and $\cY$.
\revd{Let function $Z=f(X,Y)$ take values in a finite alphabet set $\cZ$.}
$(X,Y)$ have a joint distribution $p_{XY}(x,y)$,
and their different realizations are i.i.d. In other words, $n$ consecutive
realizations $(X^n,Y^n)$ are distributed as 
$Pr(x^n,y^n)=\prod_{i=1}^{n} p_{XY}(x_i,y_i) $ for all 
$x^n = (x_1,x_2,\cdots,x_n)$ and $y^n = (y_1,y_2,\cdots,y_n)$.

The support set of $(X,Y)$ is defined as
$S_{XY}= \{ (x,y): p_{XY}(x,y)>0 \}.$
We use the notion of robust typicality \cite{Orlitsky_2001} in the following.
For $x^n \in  \cX^n $, let us denote the number of occurrences of $x \in \cX$ in $x^n$ by $N(x|x^n)$.
The set of sequences $x^n \in \cX^n$ satisfies 
\begin{align}
 \label{Eq_Typical}
 \left|\frac{1}{n} N(x|x^n) - p(x)\right| \; \leq \; \epsilon. p(x) 
\end{align}
for $\epsilon > 0 $, is called $\epsilon$-robustly typical sequences 
and is denoted by $T_{ \epsilon}^n(X)$.

\begin{definition}
A function $f(x,y)$ is called {\em component-wise one-to-one function (CWOOF)} if it satisfies the following:
\begin{enumerate}
 \item $f(x,y) \neq f(x,y') $ for all $x \in \cX, y, y'\in \cY, y \neq y'$,\\
 and
 \item $f(x,y) \neq f(x',y) $ for all $y \in \cY, x,x'\in \cX, x \neq x'$.
\end{enumerate}
This class of functions includes the binary XOR function, and in general, 
\rev{the function $a+b \mod \max(x,y)$, where $x$ and $y$ are positive integers and $ 0\leq a \leq x-1, \; 0 \leq b \leq y-1$.}
Note that computing a component-wise one-to-one function either in the broadcast network or in the relay network is equivalent to 
recovering both $X$ and $Y$ at nodes A and B.
\end{definition}

\subsection{Zero-error function computation}
\label{Model_zero_error} 
\revd{\noindent \underline{Relay Network:}} On observing $X^n$ and $Y^n$ respectively, A and B send messages
$M_A$ and $M_B$ using prefix free codes such that
$E|M_A|=nR_A$ and $E|M_B|=nR_B$. Here $|.|$ denotes the length of the
respective message in bits.
C then broadcasts a message $M_C$ with $E|M_C|=nR_C$
to A and B. Each of A and B then decode $f(X_i,Y_i);\;i=1,2,\cdots,n$ from
the information available to them.
For the relay network, a  $(2^{nR_A},2^{nR_B},2^{nR_C},n)$ 
variable length scheme consists of three encoders
$$ \phi_{A}: \cX^{n} \longrightarrow \{ 0,1\}^{*}, \quad
\phi_{B}: \cY^{n} \longrightarrow \{ 0,1\}^{*}, \quad \phi_{C}:\phi_{A}(\cX^{n}) \times \phi_{B}(\cY^{n}) \longrightarrow \{ 0,1\}^{*},$$
and two decoders
\begin{align}
\psi_A: & \cX^n\times \phi_{C}\left(\phi_{A}(\cX^{n}) \times \phi_{B}(\cY^{n})\right) \longrightarrow \cZ^n, \label{Eq_zero_dec1}\\
\psi_B: & \cY^n \times \phi_{C}\left(\phi_{A}(\cX^{n}) \times \phi_{B}(\cY^{n})\right)  \longrightarrow \cZ^n \label{Eq_zero_dec2}.
\end{align}
\label{Pg_Def_Dec}
Here $\{ 0,1\}^{*}$ denotes the set of all finite length binary sequences.
Let us define $\hat{Z}_A^n = \psi_A\left(X^n, \phi_C(\phi_A(X^n),\phi_B (Y^n)) \right) $
and $\hat{Z}_B^n = \psi_B(Y^n,\phi_C(\phi_A(X^n),\phi_B (Y^n)))$ to be
the decoder outputs.
The probability of error for a $n$ length scheme is defined as 
\begin{align}
 P_e^{(n)} \triangleq Pr & \{  (\hat{Z}_A^n,\hat{Z}_B^n)  \neq (Z^n,Z^n)\}. \label{Prob_err}
\end{align}

The rate triple $(R_{A}, R_{B}, R_{C})$ of a code is defined as
\begin{eqnarray*}
R_{A} & = & \frac{1}{n} \sum_{x^{n}} Pr(x^{n}) \mid \phi_{A}(x^{n}) \mid \\
R_{B} & = & \frac{1}{n} \sum_{y^{n}} Pr(\revd{y^{n}}) \mid \phi_{B}(y^{n}) \mid \\
R_{C} & = & \frac{1}{n} \sum_{(x^{n},y^{n})} Pr(x^{n},y^{n}) \mid \phi_{C}(\phi_{A}(x^{n}), \phi_{B}(y^{n})) \mid.
\end{eqnarray*}
\label{Pg_Def_Rb}

A rate triple $(R_{A}, R_{B}, R_{C})$ is said to be achievable 
with zero-error if for any $\epsilon>0$, there exists a scheme
\revd{with $P_e^{(n)} =0$} for a large enough $n$ such that 
$\frac{1}{n} E|M_A| \leq R_A + \epsilon,\frac{1}{n} E|M_B| \leq R_B + \epsilon $
and $\frac{1}{n} E|M_C| \leq R_C + \epsilon$. 
The rate region $\RNzerre$ is the
closure of the convex hull of all achievable rate triples.
The above setup is known as 
restricted input setup in the literature.

We now define the function computation 
in the relay network under a stricter setting, known as the
unrestricted input setup. 
A $(2^{nR_A},2^{nR_B},2^{nR_C},n)$ code for unrestricted input setup
consists of three encoders and two decoders which 
are defined as before.
Let $(\psi_A(\cdot))_i$ and 
$(\psi_B(\cdot))_i$ denote
the $i$-th components of $\psi_A(\cdot)$ and $\psi_B(\cdot)$ respectively.
A scheme is called a {\it unrestricted input scheme}
if for each $x^n\in \cX^n, y^n\in \cY^n$, and $i=1,2,\cdots,n$,
$$(\psi_A(x^n,\phi_C(\phi_A(x^n),\phi_B(y^n))))_i = f(x_i,y_i)$$
and
$$ (\psi_B(y^n,\phi_C(\phi_A(x^n),\phi_B(y^n))))_i = f(x_i,y_i)$$
if $(x_i,y_i) \in S_{XY}$. Note that this is a stricter condition than
$P_e^{(n)}=0$. A pair of vectors $(x^n, y^n)$ for which a component
$(x_i, y_i)$ is outside the support set $S_{XY}$, does not contribute
to $P_e^{(n)}$, and thus in the original zero-error problem setup, 
the decoders are also not required to correctly compute the other components.
However, the unrestricted setup requires the decoders to compute
the function correctly on all the components where $(x_i,y_i) \in S_{XY}$.
Achievable rates and the rate region $\RNzerun$
under the unrestricted setup are defined similarly as before.

\revd{\noindent \underline{Broadcast Function Network:}}
For the broadcast function network shown in Fig.~\ref{Broadcast_network},
a variable length code for the function computation problem consists of one encoder
\begin{align*}
\phi_C: & \cX^{n} \times \cY^{n} \longrightarrow \{0,1\}^*,
\end{align*}
and two decoders
\begin{align}
 \psi_A: & \phi_C (\cX^n \times \cY^{n})  \times \cX^{n} \longrightarrow \cZ^n, \label{Eq_Dec1} \\
 \psi_B: &  \phi_C (\cX^n \times \cY^{n}) \times \cY^{n}  \longrightarrow \cZ^n. \label{Eq_Dec2}
\end{align}
The rate of a code is defined as 
$\frac{1}{n} \sum_{(x^n,y^n)}Pr(x^n,y^n)|\phi_C(x^n,y^n)|$, and the outputs
of the decoders are given by $\hat{Z}_A^n = \psi_A\left(X^n, \phi_C(X^n,Y^n) \right) $
and $\hat{Z}_B^n = \psi_B(Y^n,\phi_C(X^n,Y^n))$.
A rate $R$ is said to be achievable with zero-error if \revd{for any $\epsilon>0$, there is a 
code of some length $n$ with rate $R+\epsilon$} and 
$P_e^{(n)} \triangleq Pr \{  (\hat{Z}_A^n,\hat{Z}_B^n)  \neq (Z^n,Z^n)\}=0$.
\revd{The optimal zero-error rate  $\BFNzer$ is defined as
the infimum of the set of all achievable rates. Note that $\BFNzer$ is the optimal rate under restricted input setup.}
\label{Pg_rateR}

\subsection{$\epsilon$-error function computation}
\label{Model_epsl_error}

\revd{\noindent \underline{Relay Network:}}
A fixed length $(2^{nR_A},2^{nR_B},2^{nR_C},n)$ code for function computation 
in the relay network consists of three encoder maps
\begin{align*}
\phi_{A}: &\cX^{n} \longrightarrow \{1,2,\cdots, 2^{nR_A}\},\\
\phi_{B}: & \cY^{n} \longrightarrow \{1,2,\cdots, 2^{nR_B}\},\\
\phi_{C}: & \phi_{A}(\cX^{n}) \times \phi_{B}(\cY^{n}) \longrightarrow \{1,2,\cdots, 2^{nR_C}\}
\end{align*}
and two decoder maps as defined in \eqref{Eq_zero_dec1}, \eqref{Eq_zero_dec2}.
%
%
%
A rate triple $(R_A,R_B,R_C)$ is said to be achievable with $\epsilon$-error if there exists a sequence of $(2^{nR_A},2^{nR_B},2^{nR_C},n)$ codes
such that probability of error $P_e^{(n)} \rightarrow 0$ as $n \rightarrow \infty$. The achievable rate region $\RNeps$ is the 
closure of the convex hull of all achievable rate triples.

\revd{\noindent \underline{Broadcast Function Network:}}
For the broadcast function network, a $(2^{nR},n)$ code consists of one encoder map
\begin{align*}
\phi_C: & \cX^{n} \times \cY^{n} \longrightarrow \{1,2,\cdots, 2^{nR}\}
\end{align*}
and the two decoder maps as defined in \eqref{Eq_Dec1}, \eqref{Eq_Dec2}. 
%
A rate $R$ is said to be achievable with $\epsilon$-error if there exists a sequence of $(2^{nR},n)$ codes
for which $P_e^{(n)} \rightarrow 0 $ as $n\rightarrow \infty$.
The optimal broadcast rate $\BFNeps$ in this case is
the infimum of the set of all achievable rates.

\begin{table}[h]
\centering
\revd{
\begin{tabular}{ |p{4cm}|p{8cm}|p{4cm}|} 
 \hline
 & \thead{Zero-error} & \thead{ $\epsilon$-error}\\
 \hline
 BFN-CSI - optimal rates & $\bullet$ $\BFNzer$ 
   & $\bullet$ $\BFNeps$\\
 \hline
 RN - rate regions& $\bullet$ $\RNzerre$ & $\bullet$ $\RNeps$  \\
 & $\bullet$ $\RNzerun$ - For unrestricted i/p setup &  \\
 \hline 
\end{tabular}
} 
\caption{Notations for different rate regions}\label{Tab_Notation_Region} 
\end{table}

\subsection{Graph theoretic definitions}
\label{Model_graph_definitions}

Let $G$ be a graph with vertex set $V(G)$ and edge set $E(G)$.
\revd{For two graphs $G_1$ and $G_2$ with $V(G_1)\cap V(G_2) = \emptyset$, {\it union graph} $G_1 \cup G_2$ is defined as the graph with
vertex set $V(G_1)\cup V(G_2)$ and edge set $E(G_1)\cup E(G_2)$. If $V(G_1) = V(G_2)$, then the union graph is defined
to be the graph with vertex set $V(G_1)$ and edge set $E(G_1)\cup E(G_2)$.}
A set $I \subseteq V(G) $ is called an
 {\it independent set} if no two vertices in $I$ are adjacent in $G$.
Let $\Gamma(G)$ denote the set of all independent sets of $G$.
A clique of a graph $G$ is a complete subgraph of $G$.
A clique of the largest size is called a maximum clique.
The number of vertices in a maximum clique is called  clique number of $G$ and is denoted by $\omega(G)$.
The chromatic number of $G$, denoted by $\chi(G)$, is the minimum number of colors required 
to color the graph $G$. A graph $G$ is said to be  {\it perfect} if for any vertex induced
subgraph $G'$ of $G$, $\omega(G') = \chi(G') $.
Note that the vertex disjoint union of perfect graphs is also perfect.

The $n$-fold OR product of $G$, denoted by $G^{\vee n}$,
is defined by $V(G^{\vee n}) = (V(G))^n$ and $E(G^{\vee n})=
\{(v^n,v'^n): (v_i,v'_i)\in E(G) \mbox{ for some } i\}$.
The $n$-fold AND product of $G$, denoted by $G^{\wedge n}$,
is defined by $V(G^{\wedge n}) = (V(G))^n$ and  
$E(G^{\wedge n})=
\{(v^n,v'^n): v^n \neq v'^n,\mbox{ and either } v_i=v'_i \mbox{ or } (v_i,v'_i)\in E(G) \mbox{ for all } i\}$.

For a graph $G$ and a random variable $X$ taking values in $V(G)$,
$(G,X)$ represents a {\it probabilistic graph}.
Chromatic entropy~\cite{Alon_1996} of $(G,X)$ is defined as
\begin{align*}
 H_{\chi}(G,X) &= \mbox{min} \{H(c(X)): \: c  \mbox{ is a coloring of } G \}.
\end{align*}
\label{Pg_Power_set}
Let $W$ be distributed over the \revd{power set of $\cX$}. 
The graph entropy~\cite{Korner_1973,Simonyi_1995} of the probabilistic graph $(G,X)$ is defined as
\begin{align}
 H_G(X) = \min_{X\in W \in \Gamma(G)} I(W;X),
 \label{eq:gentropy}
\end{align}
where $\Gamma(G)$ is the set of all independent sets of $G$. Here the minimum is taken over all 
conditional distributions $p_{W|X}$ which are non-zero only for $X\in W$.
The following result was shown in
\cite{Alon_1996}.
\begin{align}
\lim\limits_{n\to\infty} \frac{1}{n} H_{\chi}(G^{\vee n}, X^n) = H_G(X).
\label{eq:gpentropy}
\end{align}
\label{Pg_typ_set}
\revd{ Let $T_{P_X, \epsilon }^n$ denote the $\epsilon$-typical set
of length $n$ under the distribution $P_X$, and
let $G^{\wedge n}(T_{P_X, \epsilon }^n)$ be the vertex induced subgraph of $G^{\wedge n}$ with vertex set $T_{P_X, \epsilon }^n$. }
The complementary graph entropy of $(G,X)$ is defined as 
\begin{align*}
 \bar{H}_{G}(X) = \lim\limits_{\epsilon \to 0} \limsup \limits_{n\to\infty} \frac{1}{n} \log_2 \{ \chi(G^{\wedge n}(T_{P_X, \epsilon }^n) ) \}.
\end{align*}
Unlike graph entropy, no single-letter characterization of the complementary graph entropy is known.
It was shown in \cite{Rose_2003} that
\begin{align}
\lim\limits_{n\to\infty} \frac{1}{n} H_{\chi}(G^{\wedge n}, X^n) = \bar{H}_G(X).
\label{eq:gpentropy}
\end{align}

The definition of graph entropy was extended to the conditional graph entropy in~\cite{Orlitsky_2001}.
For a pair of random variables $(X,Y)$ and for a graph $G$ defined on the support set of $X$, 
the conditional graph entropy of $X$ given $Y$ is defined as 
\begin{align}
 H_G(X|Y) = \min_{\substack{ W-X-Y\\X\in W \in \Gamma(G)}} I(W;X|Y),
 \label{eq_cond_gentropy}
\end{align}
where the minimization is over all conditional distribution $p_{W|X}$ ($= p_{W|X,Y}$) which is non-zero only for $X\in W$.

We now define some graphs suitable for addressing our problem. 
For a function $f(x,y)$ defined over $\cX \times \cY$, we define a graph called 
$f$-modified rook's graph. 
A rook's graph $G_{\cX \cY}$ over $\cX \times \cY$ is defined by the  
vertex set  $\cX \times \cY$ and edge set
$\{((x,y),(x',y')):x=x' \mbox{ or } y=y', \mbox{ but } (x,y)\neq (x',y')\}$.

\begin{definition}
For a function $f(x,y)$ the $f$-modified rook's graph 
$\frooks$ has its vertex set $\cX \times \cY $, and
two vertices $(x_{1}, y_{1})$ and $(x_{2}, y_{2})$ 
are adjacent if and only if 
i) they are adjacent in the rook's graph $G_{\cX  \cY}$ ,
ii) $(x_1,y_1), (x_2,y_2) \in S_{XY} $, and
iii)  $f(x_{1},y_{1}) \neq f(x_{2},y_{2})$.
\end{definition}

$f$-confusability graph $G_{X|Y}^f$ of $X,Y$ and $f$ was used in \cite{Shayevitz_2014, Orlitsky_2001}
to study some function computation problems.
Its vertex set is $\cX$, and two vertices $x$ and $x'$ are adjacent if and only if
$\exists \; y\in \cY$ such that
$f(x,y) \neq f(x',y)$ and $(x,y),(x',y) \in S_{XY}$. $G_{Y|X}^f$
is defined similarly. 

\begin{figure}[h]
 \begin{subfigure}[b]{0.4\textwidth}
  \centering
\includegraphics[scale =0.4]{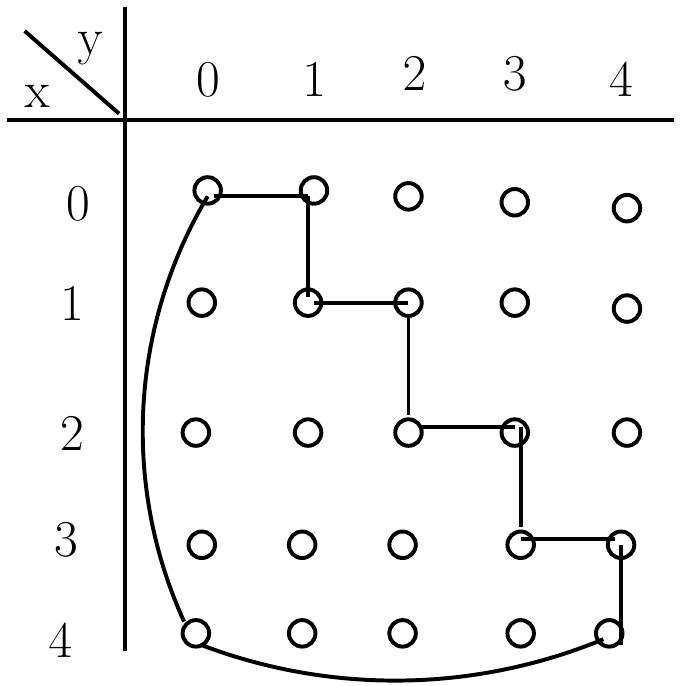}
\caption{$f$-modified rook's graph for $f(x,y)$ in \eqref{eq:function}}
\label{Rook_graph}
\end{subfigure}
\quad
 \begin{subfigure}[b]{0.5\textwidth}
   \centering
\includegraphics[scale=0.5]{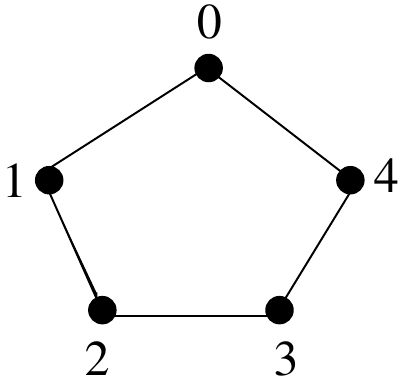}
\caption{$f$-confusability graphs $G_{X|Y}^f,G_{Y|X}^f$ for $f(x,y)$ in \eqref{eq:function}}
\label{Fig_Pentagon_Graph}
\end{subfigure}
\caption{$f$-modified rook's graph and $f$-confusability graph}
\label{Fig2}
\end{figure}

\begin{example}
\label{Ex_Zero_Pent}
Let us consider $X,Y \in \{0,1,2,3,4\}$ with distribution
\begin{equation*}
p(x,y) = \left\{
\begin{array}{cl}
  \frac{1}{10} &  \quad \mbox{if} \; y=x \mbox{ or } y=x+1 \mbox{ mod }5\\
              0  & \quad \mbox{otherwise }
\end{array} \right. ,
\end{equation*}
and the equality function
\begin{equation}
f(x,y) = \left\{
\begin{array}{cl}
 1 & \quad \mbox{if} \; x=y\\
            0 & \quad \mbox{if} \; x\neq y.
\end{array} \right. 
\label{eq:function}
\end{equation} 
The $f$-modified rook's graph for this function is shown in Fig.~\ref{Rook_graph}.
 Both $G_{X|Y}^f$ and $G_{Y|X}^f$
are the pentagon graph which is shown in Fig.~\ref{Fig_Pentagon_Graph}.
\end{example}

Next we extend the definition of $\frooks$ to $n$ instances:
\begin{definition}
\label{Def_graph_multi}
$\frooks(n)$ has its vertex set $\cX^n \times \cY^n $, and
two vertices $(x^{n}, y^{n})$ and $(x'^{n}, y'^{n})$ 
are adjacent if and only if 
\begin{enumerate}[(i)]
\item $x^n=x'^n$ or $y^n=y'^n$,
\item $Pr(x^n,y^n).Pr(x'^{n}, y'^{n}) > 0$,
\item $\exists$ an $i \in$\revd{$\{1,\cdots,n\}$} such that $f(x_{i},y_{i}) \neq f(x'_{i},y'_{i})$.
\end{enumerate}
\end{definition}
\label{Pg_braces}

To address the unrestricted input setup,
we define the following graph for $n$ instances.
\begin{definition}
\label{Def_graph_multi}
$\frooksun(n)$ has its vertex set $\cX^n \times \cY^n $, and
two vertices $(x^{n}, y^{n})$ and $(x'^{n}, y'^{n})$ 
are adjacent if and only if 
\begin{enumerate}[(i)]
\item $x^n=x'^n$ or $y^n=y'^n$,
\item   $\exists$ an $i \in$\revd{$\{1,\cdots,n\}$} such that $f(x_{i},y_{i}) \neq f(x'_{i},y'_{i})$ and
        $(x_{i},y_{i}), (x'_{i},y'_{i}) \in S_{XY}$.
\end{enumerate}
\end{definition}
It is easy to see that the graph $\frooks(n)$ is a subgraph of $\frooksun(n)$. 
Note that for $n=1$, these two graphs are the same.

\comment{
To get an inner bound for the rate region, we address a more constrained version
of the problem 
called  the OR product setup. 
In this setup, for $n$ instances we  consider the $n$-fold OR product of the
$f$-modified rook's graph. From the definition of the OR product graph,
two nodes $(x^n,y^n)$ and $(x'^n,y'^n)$ are adjacent
if and only if $(x_i,y_i)$ and $(x'_i,y'_i)$ are adjacent in $\frooks$ for some $i$.
It is easy to see that the $n$ instance graph $\frooks(n)$ is a subgraph of $(\frooks)^{\vee n}$. 
To address the  OR product setup, we give some definitions as follows.
}

\revd{Consider a graph $G$ with vertex set  $\cV$, where $\cV$ has a Cartesian representation given 
by a one-to-one mapping $\pi: \cV \rightarrow \cX \times \cY$.
For such a graph, the chromatic entropy region was defined in \cite{Shayevitz_2014}  as follows. 
If $c_1$ and $c_2$ are two maps of $\cX$ and $\cY$ into $\{0,1\}^*$ respectively, then
$c_1\times c_2$ denotes the map given by $(c_1\times c_2)(x,y) = 
(c_1(x),c_2(y))$.
A triple $(c_1,c_2,c)$ of functions of
respectively $\cX, \cY, \cV$ into $\{0,1\}^*$ is called a {\it color cover} 
for $G$ if
\begin{enumerate}[i)]
 \item $(c_1 \times c_2) \circ \pi $ and $c$ are colorings of $G$.
 \item $c_1 \times c_2$ is a refinement of $c$, i.e., $\exists$ a
mapping $\theta:(c_1\times c_2)(\cX \times \cY) \rightarrow \{0,1\}^*$ such
that $\theta \circ (c_1\times c_2) = c$.
\end{enumerate}
\label{Pg_chro_entr}
Let $\cC$ denote the set of all color covers for $G$.
For a probabilistic graph $(G,V)$, with vertex set $\cV$ having 
a Cartesian representation $\pi: \cV \rightarrow \cX \times \cY$,
let us denote $(X,Y)=\pi (V)$.
{\it Chromatic entropy region} is defined as
\begin{align*}
 H_{\chi}(G,V,\pi) & \triangleq \bigcup_{(c_1,c_2,c) \in \cC}\{(b_1,b_2,b): b_1 \geq H(c_1(X)), b_2 \geq H(c_2(Y)), b \geq H(c(V))\}. 
\end{align*}
{\em Graph entropy} region was  defined in \cite{Shayevitz_2014} from 
the definition of chromatic entropy region as follows,
\begin{equation}
H(G,V,\pi) \triangleq \bigcup_n \frac{1}{n} H_{\chi}\left(G^n, V^n, \pi^n\right),\label{Eq_zero_region}
\end{equation}
where $G^n$ denotes the $n$-fold OR product graph of $G$.
}

\revd{
Let $R_{\chi}( \frooks,X,Y)$ denote the chromatic entropy region for $f$-modified rook's graph $\frooks$.
Motivated from the graph entropy region, we define the following three dimensional regions for $f$-modified rook's graph
}
\begin{equation}
Z_{X,Y}^{f} \triangleq \bigcup_n \frac{1}{n} R_{\chi}\left( \frooks(n), X^n,Y^n\right),\label{Eq_graph_region}
\end{equation}

\begin{equation}
Z_{X,Y}^{f,(u)} \triangleq \bigcup_n \frac{1}{n} R_{\chi}\left( \frooksun(n), X^n,Y^n\right).\label{Eq_unrest_region}
\end{equation}

\rev{The graph in the following definition is used to give an inner bound for the zero-error computation in the relay network (Theorem~\ref{Thm_Zero_Inner1}).}

\begin{figure}[h]
\begin{subfigure}[b]{0.3\textwidth}
 \centering 
\includegraphics[scale=0.4]{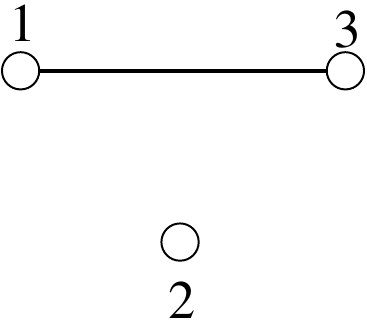}
\caption{Graphs $G_{X|Y}^f$, $G_{Y|X}^f$}
\label{Confus_Graph}
\end{subfigure} 
  \begin{subfigure}[b]{0.3\textwidth}
   \centering 
\includegraphics[scale=0.4]{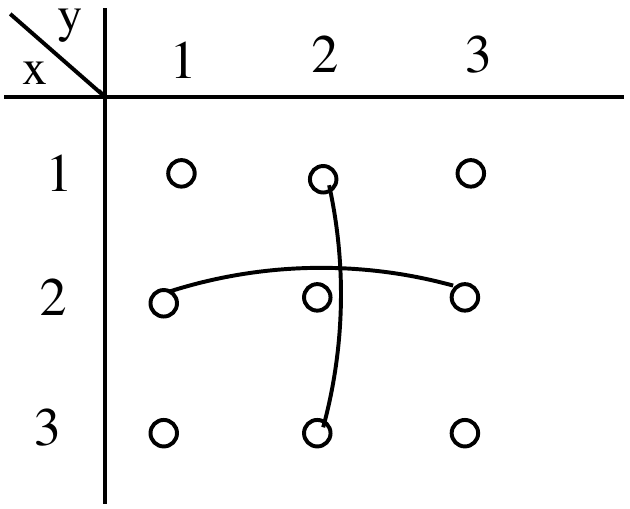}
\caption{Graph $\frooks$}
\label{Func_Orlit}
\end{subfigure}
\quad
 \begin{subfigure}[b]{0.3\textwidth}
  \centering 
 \includegraphics[scale=0.6]{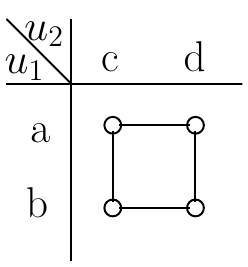}
 \caption{Graph $\faux$}
 \label{Aux_graph}
\end{subfigure}
\caption{Graphs for Example~\ref{Aux_example}}
\label{Fig2}
\end{figure}

\begin{definition}
\label{Def_aux_rooks}
 Let $U_1$ and $U_2$ be two random variables such that $X \in U_1 \in \Gamma(G_{X|Y}^f)$
 and $Y \in U_2 \in \Gamma(G_{Y|X}^f)$. 
 The random variable  $(U_1,U_2)$ over $\mathcal{U}_1 \times \mathcal{U}_2 $  has joint distribution 
 with $(X,Y)$ as $p_{X,U_1,Y,U_2}(x,u_1,y,u_2) = p(x,y)p(u_1|x)p(u_2|y)$.
We define a graph $\faux$ with vertex set $\mathcal{U}_1 \times \mathcal{U}_2$. Two vertices 
$(u_1,u_2)$ and  $(u_1',u_2')$ in $\faux$
are connected if $\exists$  $(x,y)$ and $(x',y')$ such that
\begin{enumerate}
\item  $p_{XU_1YU_2}(x,u_1,y,u_2),p_{XU_1YU_2}(x',u'_1,y',u'_2)>0 $,
 \item $x=x',u_1 =u'_1$ and $f(x,y) \neq f(x',y')$ \\
  or\\
  $y=y',u_2 =u'_2$ and $f(x,y)\neq f(x',y')$.
\end{enumerate}
\end{definition}
Note that by Definition~\ref{Def_aux_rooks}, two nodes $(u_1,u_2)$ and  $(u_1',u_2')$ are connected
in $\faux$ only if either $u_1 = u'_1$ or $u_2 = u'_2$, i.e.,
all connections are either row wise or column wise. 
Next we give an example to illustrate the above definitions.
The function in Example~\ref{Aux_example} was used in \cite{Orlitsky_2001} to explain 
the conditional graph entropy. Let us consider the same function for our function 
computation problem in the relay network.
\begin{example} \cite{Orlitsky_2001}
\label{Aux_example}
 Consider $X,Y \in \{1,2,3\}$
\begin{equation*}
 p(x,y) =\left\{
\begin{array}{cl}
  \frac{1}{6} & \quad \mbox{if} \; x\neq y\\
          0 & \quad \mbox{otherwise}
\end{array} \right.
\end{equation*}
and
\begin{equation*}
f(x,y) = \left\{
\begin{array}{cl}
 1 & \quad \mbox{if} \; x>y\\
            0 & \quad \mbox{if} \; x \leq y.
\end{array} \right. 
\end{equation*} 
 Both the confusability graphs are the same graph which is shown in Fig.~\ref{Confus_Graph}.
 The $f$-modified rook's graph for this function is shown in Fig.~\ref{Func_Orlit}.
\end{example}
In Example~\ref{Aux_example},
the distribution of $(X,Y)$ is symmetric
in $X$ and $Y$ and the function values are also symmetric.
For this example, let us consider an instance of $U_1$ and $U_2$ 
as follows. Let  $\cU_1$
be $\{\{1,2\}, \{2,3 \} \} $ and let us denote it by $\{a,b\}$ where $ a = \{1,2\} $
and $ b = \{2,3\}$. Similarly, we choose $\cU_2 $ and we denote 
it by $ \{c,d \}$, where $ c = \{1,2\} $ and $ d = \{2,3\}$. The conditional
distributions are given by $p_{U_1|X}(a|2) = p_{U_1|X}(b|2) = p_{U_2|Y}(c|2) = p_{U_2|Y}(d|2) = \frac{1}{2}$.
Now let us consider the graph $\faux$ for this function. 
The nodes $(a,c)$ and $(a,d)$ are connected in $\faux$ because $p_{XU_1YU_2}(2,a,1,c),p_{XU_1YU_2}(2,a,3,d)>0 $ and
$f(2,1) \neq f(2,3)$. By considering other pairs of nodes in $\faux$, we can verify
that the graph $\faux$ is a ``square'' graph which is 
shown in Fig.~\ref{Aux_graph}.

%
%

\section{Results}
\label{sec_characterization}

\subsection{Results for zero-error computation}
\label{Sec_zero_results}

We first give the results for the broadcast function computation problem 
shown in Fig.~\ref{Broadcast_network}.
For this problem, we show that
the optimal rate under zero-error and $\epsilon$-error
are the same.
\comment{for computing  component-wise one-to-one function
(equivalently for the complementary delivery problem)
in Fig.~\ref{Broadcast_network}.
Then we give upper and lower bounds for the  zero-error optimal rate 
for general functions.
}
Proofs of all the theorems in this 
subsection are given in Section~\ref{sec_zero_proofs}.

\comment{
\begin{theorem}[BFN-CSI]
 \label{Thm_two_Rx}
 For the broadcast function computation problem with complementary side information
(Fig.~\ref{Broadcast_network}), 
 \begin{enumerate}[(a)]
  \item \label{Thm_two_Rx_part1} 
  [\CWOO function] The optimal zero-error broadcast rate  
  for computing a component-wise one-to-one function is given by,
  \begin{align*}
    \BFNzer = \max\{ H(Y|X), H(X|Y) \}.
  \end{align*}
  \item \label{Thm_two_Rx_part2}
  [Arbitrary function]  For any function $f(X,Y)=Z$, the optimal zero-error rate $\BFNzer$ satisfies  
    \begin{align*}
       \max\{ H(Z|X), H(Z|Y) \} \leq \BFNzer \leq H_{\frooks}(X,Y).
    \end{align*}
 \end{enumerate}
\end{theorem}
}

\comment{The upper bound for $\BFNzer$ in part~\ref{Thm_two_Rx_part2} of Theorem~\ref{Thm_BFN} is obtained by 
addressing the problem under unrestricted input setup, and the given lower bound is the cutset outer bound for $\epsilon$-error computation
which also acts as an outer bound for zero-error computation.
}

\begin{theorem}
	\label{Thm_BFN}
	 For the broadcast function computation problem with complementary side information
	 \revn{shown in Fig.~\ref{Broadcast_network},} the 
	 optimal zero-error broadcast rate   $R_{(0)}^{*(BFN)}(f,X,Y )$ \revn{for computing $Z =f(X,Y)$} is given by
	 \begin{align*}
R_{(0)}^{*(BFN)}(f,X,Y ) =   \max\{ H(Z|X), H(Z|Y) \}.
	 \end{align*}
\end{theorem}
Computing a CWOOF in the relay network is equivalent to
exchanging $X$ and $Y$.
Hence using Theorem~\ref{Thm_BFN}, we get a single-letter
characterization for computing  component-wise one-to-one function in the relay network (Fig.~\ref{Relay_Network})
when the support set $S_{XY}$ is the full set.

\begin{corollary}[\CWOO in RN]
 \label{Cor_relay_XOR}
 If $S_{XY} = \cX \times \cY$, then the zero-error rate region for computing  a component-wise one-to-one function at nodes A and B in the relay network
 is given by
   \begin{align*}
 \RNzerre \triangleq & \{(R_A,R_B,R_C) : R_A \geq H(X), R_B \geq H(Y), R_C \geq  \max\{ H(Y|X), H(X|Y) \}\}.
  \end{align*}  
\end{corollary}

We note that the problem of exchanging $X$ and $Y$ through a relay has been addressed in \cite{Wyner_2002} 
under $\epsilon$-error criteria. The rate region for this problem 
under the $\epsilon$-error criteria is given by
\begin{align}
& \{(R_A,R_B,R_C) : R_A \geq H(X|Y), R_B \geq H(Y|X), R_C \geq  \max\{ H(Y|X), H(X|Y) \}\}.\label{Eq_epsl_exchng}
\end{align} 

When the sources are independent, the rate regions are clearly the same under 
$\epsilon$-error and 
zero-error criteria. When the sources are dependent with full support, 
smaller rates are possible for $R_A$ and $R_B$ under $\epsilon$-error compared to zero-error.
Even in this case, the minimum possible rate for $R_C$ is the same in both the cases.

\comment{
Next we give our results for the function computation in relay network under 
OR product setup. In Theorem~\ref{Thm_Rate_Region}, we provide 
a multi letter characterization of the rate region under this setup.
}

\begin{theorem}[RN, multiletter characterization]
 \label{Thm_Rate_Region}
  \begin{enumerate}[(a)]
  \item  \label{Thm_zero_multi}
  The zero-error rate region is given by,
  $\RNzerre =Z_{X,Y}^{f}$.
   \item  \label{Thm_unrest_multi}
   The rate region under unrestricted input setup is given by,
  $\RNzerun =Z_{X,Y}^{f,(u)}$, 
  \end{enumerate}
where $Z_{X,Y}^{f}$ and $Z_{X,Y}^{f,(u)}$ are as defined in \eqref{Eq_graph_region} and
\eqref{Eq_unrest_region} respectively.
\end{theorem}

Since a scheme under the unrestricted input setup is also 
a zero-error scheme, $\RNzerun \subseteq \RNzerre$.
The multi letter expressions for the rate regions given 
in Theorem~\ref{Thm_Rate_Region} are difficult to compute.
We give a single-letter inner bound for $\RNzerun$ in 
Theorem~\ref{Thm_Zero_Inner1}. 
\revd{
This bound is proved by considering the problem under unrestricted input setup.
Our proof technique is similar to the ones in \cite{Shayevitz_2014}.
}
\label{Pg_res_def}

\begin{theorem}[RN, zero-error inner bound]
\label{Thm_Zero_Inner1}
\revd{
\begin{enumerate}[(a)]
 \item \label{Thm_inner_part1} Let
\begin{align*}
 \cR_{I} \triangleq & \{(R_A,R_B,R_C) :  R_A  \geq I(X;U_1|Q), \; R_B  \geq I(Y;U_2|Q),\;  \\
 & \hspace{28mm} R_C \geq \min\{I(W; U_1,U_2|Q), \max\{I(X;U_1|Q), I(Y;U_2|Q)\} \} \}
  \end{align*} 
for some $p(q)p(w|u_1,u_2,q)p(u_1|x,q)p(u_2|y,q) $ such that 
\begin{enumerate}[(i)]
 \item $X \in U_1\in \Gamma(G_{X|Y}^{f})$
 \item $Y \in U_2 \in \Gamma(G_{Y|X}^{f})$
 \item $(U_1,U_2) \in W \in \Gamma(\faux) $.
\end{enumerate}
  Then $\cR_{I} \subseteq \RNzerun$.  
  \item\label{Thm_inner_part2}
  The two upper bounds for $R_C$ above, namely $I(W; U_1,U_2)$ and $ \max\{I(X;U_1), I(Y;U_2)\}$, are not comparable in general.
\end{enumerate}
}
\end{theorem}

\revd{
The proof of Theorem~\ref{Thm_Zero_Inner1} is given
in Section~\ref{Sec_Zero_Inner}.
To prove part~(\ref{Thm_inner_part2}), 
we show the following. For the function computation problem in
Example~\ref{Ex_Zero_Pent}, $\exists \; (U_1',U_2', W')$ s.t. $I(W';U_1',U_2') < \max \{ I(X;U_1), I(Y;U_2)\} $
for any $(U_1,U_2)$, and for the function computation
problem in Example~\ref{Aux_example} $\exists \; (U_1',U_2')$ s.t. $ \max \{ I(X;U_1'), I(Y;U_2')\} < I(W;U_1,U_2)$ for any $(U_1,U_2,W)$.
}

\revd{
The following corollary follows from Theorem~\ref{Thm_Zero_Inner1}.
\begin{corollary}
Any rate triple $(R_A,R_B,R_C)$ such that $$R_A \geq H_{G_{X|Y}^f}(X),\;  R_B \geq H_{G_{Y|X}^f}(Y),R_C \geq \max \{ H_{G_{X|Y}^f}(X), H_{G_{Y|X}^f}(Y)\}$$
is achievable.
\end{corollary}
}

Next we provide a sufficient condition on the joint distribution $p_{XY}$ under which 
the relay can also compute the function whenever nodes A and B compute it with zero-error.

\begin{theorem}[RN, relay's knowledge]
\label{Relay_function}
If $p(x,y)>0$ $ \forall \, (x,y) \in \cX \times \cY$, then for any 
zero-error scheme the relay can also compute the function with
zero-error.
\end{theorem}

Theorem~\ref{Relay_function} does not hold if $S_{XY} \neq \cX\times \cY$. 
We show an instance of encoding for the function given in
Example~\ref{Aux_example} to demonstrate this.
Let $\phi_A,\phi_B$ and $\phi_C$ be as follows.
\begin{equation*}
  \phi_A =\left\{
\begin{array}{cl} 
  1 & \quad \mbox{if} \; x=1\\
         0 & \quad \mbox{otherwise.}
         \end{array} \right. 
\end{equation*}

\begin{equation*}
  \phi_B  =\left\{
\begin{array}{cl} 
  1 & \quad \mbox{if} \; y=1\\
         0 & \quad \mbox{otherwise.}
\end{array} \right. 
\end{equation*}

\begin{equation*}
 \phi_C =\left\{
\begin{array}{cl} 
 
  1 & \quad \mbox{if} \quad  \phi_A=\phi_B\\
         0 & \quad \mbox{otherwise.}
 \end{array} \right.        
\end{equation*}

Here nodes A and B recover the function with zero-error, but the
relay can not reconstruct the function.
When $\phi_A = \phi_B =0$ ($(x,y)$ is either $(2,3)$ or $(3,2)$), the function
value can be both $0$ and $1$. So $H(f|\phi_A,\phi_B)>0$.

\subsection{Results for $\epsilon$-error computation}
\label{Sec_Epsl_results}

In this section, we give our results for $\epsilon$-error function 
computation in the relay network (RN).
\label{Pg_Eps_err}
\rev{
Using Lemma~\ref{Clm_rate_eqv} given in the appendix, 
we can observe that in RN, if a rate triple $(R_A,R_B,R_C)$ is achievable 
under zero-error, then  $(R_A+\delta,R_B+\delta,R_C+\delta)$ is achievable under $\epsilon$-error for any $\delta >0$.
This shows that  in general
 the rate region for computing a function in RN with $\epsilon$-error is equal to or
larger than the rate region for computing the 
function with zero-error. }
\comment{
Computing with zero error is a stricter requirement than computing
with $\epsilon$ error, and 
any zero-error code is also an $\epsilon$-error code. So in general
the rate region for computing a function in RN with $\epsilon$-error is 
larger than the rate region for computing the 
function with zero-error. }
In Example~\ref{Ex_XOR1}, we give an instance for which the rate region under $\epsilon$-error
is strictly larger than the rate region under zero-error.
Proofs of all the theorems in this 
subsection are given in Section~\ref{Sec_asym_bounds}.

\begin{example}
\label{Ex_XOR1}
Let us consider computing $X \oplus Y $ for
a {\it doubly symmetric binary source} (DSBS($p$))
$(X,Y)$ where $p_{X,Y}(0,0)= p_{X,Y}(1,1)= (1-p)/2$ and $p_{X,Y}(0,1)= p_{X,Y}(1,0)=p/2$.
From Corollary~\ref{Cor_relay_XOR},
we have the zero-error rate region as
$ \{(R_A,R_B,R_C) : R_A \geq  1, R_B \geq 1, R_C \geq H(p) \}$.
As noted before, computing $X \oplus Y $ in the relay network is same as exchanging $X$ and $Y$.
The $\epsilon$-error rate region for exchanging $X$ and $Y$ through the relay
is given in \eqref{Eq_epsl_exchng}. Computing this for 
DSBS($p$) $(X,Y)$ gives the rate region as
$ \{(R_A,R_B,R_C) : R_A, R_B, R_C \geq H(p) \} $.
\end{example}

For arbitrary functions, we do not have a single-letter characterization for
the $\epsilon$-error rate region. Next lemma gives a cutset outer bound for 
the $\epsilon$-error rate region.
\label{Pg_Colon}
\begin{lemma}
\label{Lem_Asym_Outer_Bound}
 \begin{enumerate}[(a)]
  \item   \label{Lem_Asym_Outer1} 
  [Cutset outer bound]
 Any achievable rate triple $(R_{A}, R_{B}, R_{C}) \in \RNeps$ for RN satisfies the following\revd{:}
 \begin{align}
  R_A & \geq H_{G_{X|Y}^{f}}(X|Y), \quad R_B  \geq H_{G_{Y|X}^{f}}(Y|X), \quad R_C  \geq \max\{H(Z|X), H(Z|Y)\}. \label{Eq_Asym_Outer}
 \end{align}
  \item   \label{Lem_Asym_Outer2} 
  Equality in \eqref{Eq_Asym_Outer} can be achieved individually for either $(R_A,R_B)$ or $R_C$.
 \end{enumerate} 
\end{lemma}

\begin{remark}
\label{rem_asym_outer} 
We suspect the cutset bound to be loose, though we do not 
have an example to show this.
For all the example functions where we have a single-letter 
characterization of the rate region, the cutset
outer bound in \eqref{Eq_Asym_Outer} is seen to be tight. 
Example~\ref{Ex_tight_outer} provides a class of functions for which
the cutset outer bound is tight.
\end{remark}

Next we propose two achievable schemes  for
the $\epsilon$-error computation problem.
These two schemes are the extensions of the zero-error schemes given in Theorem~\ref{Thm_Zero_Inner1}.

\begin{theorem}[RN, $\epsilon$-error inner bound]
\label{Thm_Epsilon_Inner1}
\begin{enumerate}[(a)]
 \item \label{Thm_asym_inner_part1} Let
\begin{align*}
  \eR_{I1}^{\epsilon } \triangleq & \{(R_A,R_B,R_C) :    R_A  \geq I(X;U_1|U_2,Q), R_B  \geq I(Y;U_2|U_1, Q),\\
  & \; R_A + R_B  \geq I(X,Y; U_1, U_2|Q), R_C  \geq \max\{ I(W; U_1|U_2,Y,Q), I(W;U_2|U_1,X,Q) \} \}
  \end{align*} 
for some $p(q)p(w|u_1,u_2,q)p(u_1|x,q)p(u_2|y,q) $ such that 
\begin{enumerate}[(i)]
 \item $X \in U_1\in \Gamma(G_{X|Y}^{f})$
 \item $ Y \in U_2 \in \Gamma(G_{Y|X}^{f})$
 \item $(U_1,U_2) \in W \in \Gamma(\faux) $.
\end{enumerate}
 \begin{align*}
\hspace{-3cm} \mbox{Let} \hspace{3cm}  \eR_{I2}^{\epsilon} \triangleq & \{(R_A,R_B,R_C) : R_A \geq  H_{G_{X|Y}^{f}}(X|Y), R_B \geq H_{G_{Y|X}^{f}}(Y|X),\\
 & \hspace{28mm} R_C  \geq \max\{H_{G_{X|Y}^{f}}(X|Y), H_{G_{Y|X}^{f}}(Y|X)\} \}.
 \end{align*}
  Let $\eR_{I}^{\epsilon}$ be the convex closure of $\eR_{I1}^{\epsilon} \cup \eR_{I2}^{\epsilon}$. Then $\eR_{I}^{\epsilon} \subseteq \RNeps$.
  
  \item\label{Thm_asym_inner_part2}
  Neither of $\eR_{I1}^{\epsilon}$ and  $\eR_{I2}^{\epsilon}$ is a subset of 
  the other in general.
\end{enumerate}
\end{theorem}

The proof of Theorem~\ref{Thm_Epsilon_Inner1} is given in 
Section~\ref{Asym_Inner}.
To prove part~(\ref{Thm_asym_inner_part2}), 
we show that for computing AND for a $DSBS(p)$ source,
the  rate triple \revd{$(H(p),H(p),H(p)) \in \eR_{I2}^{\epsilon} \setminus \eR_{I1}^{\epsilon}$}, and
\revd{$(1,H(p),\frac{1}{2} H(p)) \in \eR_{I1}^{\epsilon} \setminus \eR_{I2}^{\epsilon}$}.

\begin{example}
\label{Ex_tight_outer} 
Let us consider the functions where one of the confusability graphs is empty. 
W.l.o.g., let us assume that $G_{Y|X}^{f}$ is empty. Then on the support
set $S_{XY}$, the function $f$ can be computed from $X$ alone.
This implies that node A 
can compute the function with zero-error from $X$, and $H_{G_{Y|X}^{f}}(Y|X) = 0$.
Let us consider $H_{G_{X|Y}^{f}}(X|Y)$.  In general, $H_{G_{X|Y}^{f}}(X|Y) \geq H(Z|Y)$.
For a given $Z=z$, let us consider the set of all $x$, $A_z=\{x: f(x,y)=z, \mbox{ for some }y \mbox{ s.t. } (x,y)\in S_{XY} \}$.
Since here for $X=x$, $f(x,y') = f(x,y'')$ for any $(x,y'),(x,y'') \in S_{XY}$,
$A_z$ is an independent set of $G_{X|Y}^{f}$. 
Let $\cA$ denote the set of all $A_z$, and
$W=A_Z$.
Since $Z$ is a function of $X$, we have $W=g(X)$ for some function $g$.
This $W$ in \eqref{eq_cond_gentropy} gives that $I(W;X|Y) = H(Z|Y)$.
So we get $H_{G_{X|Y}^{f}}(X|Y) = H(Z|Y)$.
Then we get $ \eR_{I2}^{\epsilon}$ in Theorem~\ref{Thm_Epsilon_Inner1}
as $ \{(R_A,R_B,R_C) : R_A \geq  H(Z|Y), R_B \geq 0,
R_C  \geq H(Z|Y)\}$. It is easy to check that the cutset outer bound in \eqref{Eq_Asym_Outer} 
also gives the same rate region. This shows that for functions where one of the 
confusability graph is empty, the cutset outer bound is tight. 
\end{example}

\begin{theorem}
\label{Thm_Rooks}
Let $f_1,f_2$ be two functions of $(X,Y)$.
\begin{enumerate}[(a)]
 \item \label{Thm_rooks_part1} 
 If $E(\frookA) \subseteq E(\frookB)$, then $(i)$ $\RNzerunA \supseteq  \RNzerunB$, 
 $(ii)$ $\RNepsA \supseteq  \RNepsB$.
  \item \label{Thm_rooks_part2}   
 If $\frookA$ is isomorphic to $\frookB$, then $(i)$  $\RNzerunA = \RNzerunB$,
  $(ii)$ $\RNepsA = \RNepsB$.
\end{enumerate}
\end{theorem}

For any arbitrary function $f$ of $(X,Y)$, if $\frooks$ is isomorphic to
 the the $f$-modified rook's graph  for exchanging $X$ and $Y$
(i.e. computing a \revd{component-wise one-to-one function}), then the rate region $ \RNeps$ is given by
\eqref{Eq_epsl_exchng}.  $\RNepsA =\RNepsB$ does not imply
the isomorphism between $\frookA$ and $\frookB$. We show this through the following example.

\begin{example}
\label{Ex_Graph}
For a DSBS($p$)  $(X,Y)$, let functions $f_1,f_2$ of $(X,Y)$ be defined as\footnote{Here + is sum, not XOR. 
In particular, $f_1(1,1) = 2$.} $f_1 = X+Y$ 
and $f_2 = Y\cdot(X+Y)$.
For these functions, $\frookA$ and
$\frookB$ are shown in Fig.~\ref{Fig2}.
The graph $\frookA$ is  same as  \revd{ as the $f$-modified rook's graph for computing a component-wise one-to-one function}.
Using Theorem~\ref{Thm_Rooks}, we get 
$\RNepsA = \{(R_A,R_B,R_C) : R_A, R_B, R_C \geq H(p) \}$.
For function $f_2$, since 
graphs  $G_{X|Y}^{f_2}$ and  $G_{Y|X}^{f_2}$ are complete graphs,
$H_{G_{X|Y}^{f_2}}(X|Y) = H(X|Y)$, and $H_{G_{Y|X}^{f_2}}(Y|X) = H(Y|X)$.
Further, we have
$H(Z_2|X)= H(p)$ and $H(Z_2|Y)= \frac{1}{2} H(p)$.
This implies that $ \max\{H_{G_{X|Y}^{f}}(X|Y), H_{G_{Y|X}^{f}}(Y|X)\} =   \max\{H(Z|X), H(Z|Y)\} = H(p)$.
Then the region given by $\eR_{I2}^{\epsilon}$ 
in Theorem~\ref{Thm_Epsilon_Inner1} is same as the region given by the  cutset outer bound in \eqref{Eq_Asym_Outer}.
So we get
$\RNepsB = \{(R_A,R_B,R_C) : R_A, R_B, R_C \geq H(p) \}$ which
is same as $\RNepsA$. 
Here, even though $\RNepsA = \RNepsB$,
$\frookA$ is not isomorphic to $\frookB$.
\end{example}

\begin{figure}[h]
 \begin{subfigure}[b]{0.4\textwidth}
 \centering
\includegraphics[scale =0.6]{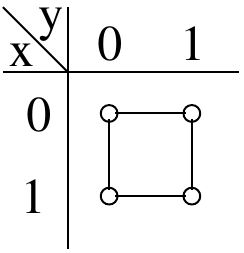}
\caption{$\frookA$}
\label{Modified}
\end{subfigure}
 \begin{subfigure}[b]{0.4\textwidth}
 \centering
\includegraphics[scale=0.6]{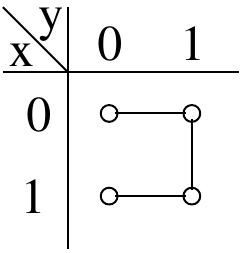}
\caption{$\frookB$}
\label{Pentagon_Graph}
\end{subfigure}
\caption{ Graphs $\frookA$ and $\frookB$ in Example~\ref{Ex_Graph}}
\label{Fig2}
\end{figure}

In \cite{Han_1987}, Han and Kobayashi considered the function computation
problem where two encoders encode $X^n$ and $Y^n$, and a decoder wants
to compute $f(X,Y)$ from the encoded messages.
They gave necessary and sufficient conditions under which
the function computation rate region coincides with the 
Slepian-Wolf region. The conditions were based on a probability-free structure 
of the function $f(X,Y)$, assuming that $S_{XY}=\cX\times\cY$. 
\revd{
For our function computation problem, 
in general, if $\frooks$ is not the same as the $f$-modified rook's graph for a component-wise one-to-one function, then
the equality $\RNeps = \RNepsX$ }
also depends on $p_{XY}$ even when $S_{XY}=\cX\times\cY$.
In particular, for the function $f_2$  in Example~\ref{Ex_Graph},
the equality $\RNeps = \RNepsX$ depends on 
the distribution $p_{XY}$. This is illustrated in Example~\ref{Ex_Han_Kob}.
Thus we observe that the characterization of $\RNeps 
= \RNepsX$ in the relay network
cannot have a probability-free structure.

\begin{example}
 \label{Ex_Han_Kob}
 Let us consider the function $f_2$ in Example~\ref{Ex_Graph}.
 When $p_{XY}$ is DSBS($p$), it is shown in Example~\ref{Ex_Graph} that
 $\RNeps = \RNepsX$ . Let us consider the same function for the 
following distribution
\begin{align*}
  p(0,0) &= p(1,0) = \frac{1}{6},\\
  p(0,1) &= p(1,1) = \frac{1}{3}.
\end{align*}
We have
 $H(X|Y) = H(X) = 1$ and $ H(Y|X) = H(Y) = H(\frac{1}{3})$.
So we get 
$\RNepsX =  \{(R_A,R_B,R_C) : R_A \geq 1 , R_B \geq H(\frac{1}{3}), R_C \geq 1 \}$. For $Z=f_2(X,Y)$,
$ H(Z|Y)  = \frac{2}{3}$, and  $H(Z|X)  = H(1/3) \approx 0.91$.
Let us consider an instance of encoding where A and B
communicate $X^n$ and $Y^n$ to the relay with rates $R_A=H(X)$ and $R_B=H(Y)$
respectively; and the relay computes $Z^n$ and 
use Slepian-Wolf binning to compress it at a rate
$R_C =\max \{H(Z|X), H(Z|Y)\}$.
Then the function computation at A and B follows from the Slepian-Wolf decoding.
For this scheme, the rate triple $(1,H(1/3), H(1/3))$ is achievable.
Clearly, $(1,H(1/3), H(1/3)) \notin \RNepsX$
and we get  $\RNeps \neq \RNepsX$.
\end{example}

%
\section{Zero error computation: Proofs of Theorems~\ref{Thm_BFN}-~\ref{Relay_function} }
\label{sec_zero_proofs}

\rev{
	\subsection{ Proof of  Theorem~\ref{Thm_BFN}}
	The optimal $\epsilon$-error rate $R_{(\epsilon)}^{*(BFN)}(f,X,Y)$  is given by $\max\{ H(Z|X), H(Z|Y) \}$ which follows from the Slepian-Wolf result~\cite{Slepian_1973}.
	Using Lemma~\ref{Clm_rate_eqv}, we can observe that $R_{(\epsilon)}^{*(BFN)}(f,X,Y) \leq R_{(0)}^{*(BFN)}(f,X,Y )$.
	Next we show that $R_{(0)}^{*(BFN)}(f,X,Y) \leq \max\{ H(Z|X), H(Z|Y) \}$.
The code has two constituents: a subset $S\subseteq\cX^n\times\cY^n$, and
a random binning of all sequences $z^n$ into $2^{R'n}$ bins, 
        where  $R' = \max \{H(Z|Y),H(Z|X)\}+\frac{\delta}{2}$.
	Let $S$ be the set of all $(x^n,y^n)$ satisfying at least one of the following:\\
	E1:  $(x^n,f(x^n,y^n))\notin T_{ \epsilon}^n(XZ)$,\\
	E2: $(y^n,f(x^n,y^n))\notin T_{ \epsilon}^n(YZ)$,\\
	E3:  $\exists \; z'^n \neq f(x^n,y^n) $ such that $(z'^n,x^n) \in  T_{ \epsilon}^n(ZX) $, and it is in the same bin as $f(x^n,y^n)$,\\
	E4: $\exists \; z'^n \neq f(x^n,y^n) $ such that $(z'^n,y^n) \in T_{ \epsilon}^n(ZY) $, and it is in the same bin as $f(x^n,y^n)$.\\
The sequences in $S$ are indexed by a fixed length code of length at most
$n(\log |\cX| + \log |\cY|)$. The overall code consists of the indices of
$S$ and the indices of the bins, distinguished by an additional prefix bit.
}

\rev{
	The encoder sends the bin index of $f(X^n,Y^n)$ if $(X^n,Y^n) \in S^{\complement}$. Otherwise, it sends the index of $(X^n,Y^n)$ in $S$. 
	If node A receives a bin index, then it finds the unique $Z^n$ which is jointly typical with $X^n$. 
	Otherwise, node A gets to know $(X^n,Y^n)$ from its index in $S$, and computes $Z^n=f(X^n,Y^n)$.
	Node B  follows similar decoding. There is  no decoding error 
	either for node A or B under this scheme, as all sequences $(x^n,y^n)$
which could have resulted in a decoding error are separately transmitted
using their index in $S$.
	From the Slepian-Wolf result~\cite{Slepian_1973}, we know that 
   the probability  $Pr(E1\cup E3)$  is less than or equal to $2^{-n\delta/2}$ for large enough $n$.
   Similarly, $Pr(E2 \cup E4)$ is less than or equal to $2^{-n\delta/2}$.
   Thus by union bound, $Pr(S)\leq Pr(E1\cup E3)+Pr(E2\cup E4)\leq 2\times2^{-n\delta/2}$.
	Since $\log |S|$ is linear in $n$, the overall average length of the code is at most
\begin{align*}
& Pr(S)n(\log |\cX| + \log |\cY|) + Pr(S^\complement)nR'\\
& \leq 2\times2^{-n\delta/2}n(\log |\cX| + \log |\cY|)+n\left(\max \{H(Z|Y),H(Z|X)\}+\frac{\delta}{2}\right)\\
& \leq n\frac{\delta}{2}+n\left(\max \{H(Z|Y),H(Z|X)\}+\frac{\delta}{2}\right)\\
& = n(\max \{H(Z|Y),H(Z|X)\} + \delta)
\end{align*}
for large enough $n$. This completes the proof of the theorem.
	\hfill{\rule{2.1mm}{2.1mm}}
}
    
{\em Proof  of Corollary ~\ref{Cor_relay_XOR}:}
First let us consider the converse for the rate region.
For $R_A$, let us consider the cut between node A and a super node consisting of B and C.
This situation arises when the relay node broadcasts the message sent by node A.
Then the problem reduces to the problem of decoding with side information studied in  \cite{Alon_1996},
where the decoder with side information $Y$ wants to recover $X$. 
\label{Pg_Alon_lemma}
\revd{Lemma~6 in \cite{Alon_1996} shows that the optimal rate is equal to $\lim\limits_{n\to\infty} \frac{1}{n} H_{\chi}(G^{\wedge n}, X^n) = \bar{H}_G(X)$.
Since the support set is full, the graph $(G,X)$ is a complete graph with vertex set $\cX$. It can be easily
verified that for a complete graph, $\bar{H}_G(X) = H(X)$.
So here we get $R_A \geq H(X)$. 
}
Similarly, $R_B \geq H(Y)$.
Now let us consider the rate $R_C$.
\revd{Any relay encoding $\phi_C(\phi_A(x^n), \phi_B(y^n))$ is also a function of $(x^n,y^n)$ and so any achieved rate $R_C$ can also be achieved if the relay
has the full information $(x^n,y^n)$. 
So the optimum $R_C$ attains its minimum value when the relay has $X$ and $Y$.}
\rev{
For a component-wise one-to-one function $Z$, $H(Z|X)=H(Y|X)$ and $H(Z|Y)=H(X|Y)$.
}
Theorem~\ref{Thm_BFN} shows that if 
relay has both $X$ and $Y$, the minimum achievable broadcast rate is 
$\max\{ H(Y|X), H(X|Y) \}$.
This completes the converse.
Now let us consider a scheme where nodes A and B communicate $X$ and $Y$ respectively
to the relay. The relay can recover $X$ and $Y$ with zero-error if $R_A > H(X)$ and $R_B > H(Y)$.
If the  relay has $X$ and $Y$, 
Theorem~\ref{Thm_BFN}
shows that the rate $  \max\{ H(Y|X), H(X|Y) \}$ is achievable for $R_C$ for computing a component-wise one-to-one function.
This proves the achievability of the rate region.
\hfill{\rule{2.1mm}{2.1mm}}

\subsection{Proof of Theorem~\ref{Thm_Rate_Region}}
\label{Sec_rate_region}
To prove Theorem~\ref{Thm_Rate_Region}, we first present some lemmas.

\begin{lemma}
 \label{Lem_zero_multi}
 For any $n\geq 1$, and given the encoding functions $\phi_A,\phi_B,\phi_C$, 
the nodes A and B can recover $f(X^n,Y^n)$ with zero-error if and only if
$\phi_C \circ (\phi_A \times \phi_B)$ is a coloring of 
$\frooks(n)$.
\end{lemma}

\begin{IEEEproof} 
Let $E(\frooks(n))$ denote the set of edges of $\frooks(n)$.
Note that 
\begin{align}
 E(\frooks(n)) & = \{((x^n,y^n),(x^n,y'^n)) \in S_{X^nY^n};  f(x_i,y_i) \neq f(x_i,y'_i)\mbox{ for some } i \} \notag  \\
& \quad \cup \{((x^n,y^n),(x'^n,y^n)) \in S_{X^nY^n};  f(x_i,y_i) \neq f(x'_i,y_i)\mbox{ for some } i \}. \label{eq:edgeset}
\end{align}
Observe that each edge is of the form $((x^n,y^n),(x^n,y'^n))$ or $((x^n,y^n),(x'^n,y^n))$.
We note that

(i) A  can recover $f(X^n,Y^n)$ with zero-error $\Leftrightarrow$
for any $(x^n,y^n),(x^n,y'^n)\in S_{X^nY^n}$ with
$f(x_i,y_i) \neq f(x_i,y'_i)$ for some $i$, $\phi_C(\phi_A(x^n),\phi_B(y^n)) \neq 
\phi_C(\phi_A(x^n),\phi_B(y'^n))$.

(ii)  B  can recover $f(X^n,Y^n)$ with zero-error $\Leftrightarrow$
for any $(x^n,y^n),(x'^n,y^n)\in S_{X^nY^n}$ with
$f(x_i,y_i) \neq f(x'_i,y_i)$ for some $i$, $\phi_C(\phi_A(x^n),\phi_B(y^n)) \neq 
\phi_C(\phi_A(x'^n),\phi_B(y^n))$.

From (i) and (ii) above, it follows that A and B can recover$f(X^n,Y^n)$ with zero-error
$\Leftrightarrow$ 
for any $((x^n,y^n)$, $(x'^n,y'^n)) \in E(\frooks(n))$, $\phi_C(\phi_A(x^n),\phi_B(y^n)) \neq
\phi_C(\phi_A(x'^n),\phi_B(y'^n))$ $\Leftrightarrow$ $\phi_C \circ (\phi_A \times \phi_B)$ is a coloring of 
$\frooks(n)$.
\end{IEEEproof}

\begin{lemma}
 \label{Lem_unrest_multi}
 For any $n\geq 1$, and given the encoding functions $\phi_A,\phi_B,\phi_C$, 
the nodes A and B can recover $f(X^n,Y^n)$ under the unrestricted input setup
if and only if $\phi_C \circ (\phi_A \times \phi_B)$ is a coloring of 
$\frooksun(n)$.
\end{lemma}

\begin{IEEEproof} 
Let $E(\frooksun(n))$ denote the set of edges of $\frooksun(n)$.
Observe that 
\begin{align}
 E(\frooksun(n)) & = \{((x^n,y^n),(x^n,y'^n)): \mbox{ for some } i  ((x_i,y_i),(x_i,y'_i)) \in E(\frooksun) \} \notag  \\
 & \quad \cup \{((x^n,y^n),(x'^n,y^n)):  \mbox{ for some } i  ((x_i,y_i),(x'_i,y_i)) \in E(\frooksun) \}.  \label{eq_unrest_edgeset}
\end{align}
We note that

(i) A can recover $f(X^n,Y^n)$ under the unrestricted input setup $\Leftrightarrow$
for any $(x^n,y^n),(x^n,y'^n)$ such that
$f(x_i,y_i) \neq f(x_i,y'_i)$ for some $i$ where $(x_i,y_i),(x_i,y'_i) \in S_{XY}$, $\phi_C(\phi_A(x^n),\phi_B(y^n)) \neq 
\phi_C(\phi_A(x^n),\phi_B(y'^n))$.

(ii) B can recover $f(X^n,Y^n)$ under the unrestricted input setup $\Leftrightarrow$ for any $(x^n,y^n),(x'^n,y^n)$ 
such that $f(x_i,y_i) \neq f(x'_i,y_i)$ for some $i$ where $(x_i,y_i),(x'_i,y_i) \in S_{XY}$,
$\phi_C(\phi_A(x^n),\phi_B(y^n)) \neq \phi_C(\phi_A(x'^n),\phi_B(y^n))$ 

From (i) and (ii) above, it follows that A and B can recover$f(X^n,Y^n)$ with zero-error
$\Leftrightarrow$ 
for any $((x^n,y^n)$, $(x'^n,y'^n)) \in E(\frooksun(n))$, $\phi_C(\phi_A(x^n),\phi_B(y^n)) \neq
\phi_C(\phi_A(x'^n),\phi_B(y'^n))$ $\Leftrightarrow$ $\phi_C \circ (\phi_A \times \phi_B)$ is a coloring of 
$\frooksun(n)$.
\end{IEEEproof}

{\em Proof of part~(\ref{Thm_zero_multi})}:
Lemma~\ref{Lem_zero_multi}
implies that for encoding functions $\phi_A,\phi_B,\phi_C$ of 
any zero-error scheme, $\phi_A, \phi_B, \phi_C\circ (\phi_A\times\phi_B)$
is a color cover for $\frooks(n)$.
Similarly, for any  color cover $(c_A,c_B,c_C)$ of 
$\frooks(n)$, let $\phi_A,\phi_B$ be any prefix-free 
encoding functions of $c_A$ and $c_B$ respectively. 
Since $c_A\times c_B$ is a refinement of $c_C$,  there exists a
mapping $\theta_C$ such that $c_C = \theta_C \circ (c_A\times c_B)$.
Taking $\phi_C$ as any prefix-free encoding of $c_C$ yields
a scheme with encoding functions $(\phi_A,\phi_B,\phi_C)$.
Thus the result follows from the definition of the region 
$Z_{X,Y}^f$. 
\hfill{\rule{2.1mm}{2.1mm}}

Proof of part~(\ref{Thm_unrest_multi}) follows along the similar lines
as that of part~(\ref{Thm_zero_multi}) using Lemma~\ref{Lem_unrest_multi}.

\subsection{Proof of Theorem~\ref{Thm_Zero_Inner1}}
\label{Sec_Zero_Inner}
We first give some lemmas which are used to prove the theorem.
\begin{lemma}
 \label{Lem_Covering}
 (Covering Lemma, \cite{Elgamal_Kim}). Let $(U,X,\hat{X}) \sim p(u,x,\hat{x})$ and
 $\epsilon' < \epsilon$. Let $(U^n,X^n) \sim p(u^n,x^n)$ be a pair of random sequences
 with $ \lim\limits_{n\rightarrow \infty} P\{ (U^n,X^n) \in  T_{\epsilon'}^n(U,X) \} = 1$,
 and let $ \hat{X}^n(m), m\in \cA$, where $|\cA| \geq 2^{nR}$, be random sequences,
 conditionally independent of each other and of $X^n$ given $U^n$, each distributed 
 according to $\prod_{i=1}^n p_{\hat{X}|U}(\hat{x}_i|u_i)$. Then, there exists
 $\delta(\epsilon)$ that tends to zero as $\epsilon \rightarrow 0 $ such that 
 $$ \lim_{n \rightarrow \infty} P\{(U^n, X^n, \hat{X}^n(m) ) \notin T_{\epsilon}^n \; \mbox{ for all } m \in \cA \} = 0,$$ 
 if $R> I(X;\hat{X}|U) +\delta(\epsilon)$.
\end{lemma}
\begin{lemma}
 \label{Lem_markov}
 (Markov Lemma, \cite{Elgamal_Kim}). Suppose that $X \rightarrow Y \rightarrow Z $ form a Markov chain.
 Let $(x^n,y^n) \in T_{\epsilon'}^n(X,Y)$, and $Z^n \sim p(z^n|y^n)$, where the conditional pmf $p(z^n|y^n)$
 satisfies the following conditions: 
 \begin{enumerate}
  \item $\lim\limits_{n \rightarrow \infty} P\{(y^n,Z^n) \in  T_{\epsilon'}^n(Y,Z)\} = 1 .$
  \item For every $z^n \in  T_{\epsilon'}^n(Z|y^n)$ and $n$ sufficiently large 
       $$2^{-n(H(Z|Y)+ \delta(\epsilon'))} \leq  p(z^n|y^n) \leq 2^{-n(H(Z|Y) - \delta(\epsilon'))} $$
       for some $\delta(\epsilon')$ that tends to zero as $\epsilon' \rightarrow 0 $.\\
 \end{enumerate}
      Then, for some sufficiently small $ \epsilon' < \epsilon $, 
      $$\lim_{n \rightarrow \infty} P\{(x^n,y^n,Z^n) \in  T_{\epsilon}^n(X,Y,Z)\} = 1 .$$   
\end{lemma}

\begin{lemma}\cite[Lemma~4]{Orlitsky_2001}
\label{Lem_zero_schem1}
There exists a function
$g$ such that $\forall (x,y)\in S_{XY}, u_2\in \Gamma(G_{Y|X}^f)$ s.t.
$y\in u_2$, $g(x,u_2) = f(x,y)$,\revd{i.e., $f(x,y)$ can be computed from $u_2$ and $y$}.
\label{Pg_funt_lemma}
%
\end{lemma}

\begin{lemma}
\label{Sche2_lemma3}
There exists functions $g_1$ and $g_2$ such that
for all $(x,y,u_1,u_2,w)\in \cX\times \cY\times\Gamma(G_{X|Y}^f)\times
\Gamma(G_{Y|X}^f)\times\Gamma(\faux)$ satisfying
$(u_1,u_2) \in w$ and $p(x,y)p(u_1|x)p(u_2|y)>0$,
$f(x,y) = g_1(x,u_1,w)=g_2(y,u_2,w)$.
\end{lemma}
\begin{IEEEproof}
For a given $X=x, U_1 = u_1$ and $W=w$, let us consider the set of possible 
$y$, $A_{x,u_1,w} = \{y':(x,y') \in S_{XY}, \mbox{ and } p(u'_2|y')>0 \mbox{ for some } u'_2 \mbox{ s.t. } (u_1,u'_2) \in w \}$.
Then we show that \\
{\bf Claim:}
$ f(x,y') = f(x,y'') $ $  \forall y',y'' \in A_{x,u_1,w}$ . 

{\it Proof of the claim:} Let us assume that for 
some $ y',y'' \in A_{x,u_1,w}$, $f(x,y') \neq f(x,y'')$.
By definition of $A_{x,u_1,w}$, $\exists u_2', u_2'' \in \Gamma (G^f_{Y|X})$,
such that $y'\in u_2', y'' \in u_2''$, and $(u_1,u_2'), (u_1, u_2'') \in w$.
But $(y',y'') \in E(G^f_{Y|X})$,  and so $ y'' \not\in u_2'$, and
thus $u_2' \neq u_2''$.
From the conditions in the lemma and the definition of $A_{x,u_1,w}$,
we have $p(x,u_1,y',u'_2), p(x,u_1,y'',u''_2)>0$.
Then by Definition~\ref{Def_aux_rooks}, $(u_1,u'_2)$ and $(u_1,u''_2)$
are connected in $\faux$. This implies that $w$ is not an independent set
of  $\faux$, which is a contradiction. This proves the claim.

Now, $g_1$ (resp. $g_2$) is defined as the unique function
value $f(x,y)$ for all $y\in A_{x,u_1,w}$ (resp. \revd{$x\in A_{y,u_2,w}$}).
\label{Pg_lemm_proof}
\end{IEEEproof}

We first give the proof of  part~(\ref{Thm_inner_part1}).

{\em Proof of part~(\ref{Thm_inner_part1})}:
In the following, we assume $\epsilon > \epsilon' > \epsilon''>0$ and $|\cQ| = 1$.
Let $\{ U_1^n(m_1) |m_1 \in\{ 1,\cdots,{2^{nR_A'}}\}\}$ be a set of independent sequences, each distributed according
to $\prod_{i=1}^n p_{U_1}(u_{1i})$.
Similarly, let  $\{ U_2^n(m_2) |m_2 \in\{ 1, \cdots,{2^{nR_B'}}\}\}$, be a set of independent sequences, each distributed according
to $\prod_{i=1}^n p_{U_2}(u_{2i})$.
Let  $\{ W^n(m_3) |m_3 \in \{ 1, \cdots, {2^{nR_C'}}\}\}$, be a set of independent sequences, each distributed according to $\prod_{i=1}^n p_{W}(w_{i})$.

\noindent \textbf{ Encoding at node A}:
\label{Pg_foot}

For a given $x^n$, node A chooses an index $m_1$ (if any)
such that $(x^n, U_1^n(m_1)) \in T_{\epsilon''}^n(X,U_1)$.
The encoding at node A is given by\footnote{\revd{Transmission of an $x^n$ sequence is done by first converting the sequence to a binary sequence of maximum length $n \log |\cX|$ bits. 
An extra prefix bit is added to distinguish the sequences $x^n$ and $m_1$. Similar operation is done for a $y^n$ sequence too.}}
\begin{align*}
 \phi_A(x^n) & = \begin{cases}
               m_1 \; & (x^n, U_1^n(m_1)) \in T_{\epsilon''}^n(X,U_1) \\
               x^n \; & \mbox{if } (x^n, U_1^n(m_1)) \not\in T_{\epsilon''}^n(X,U_1) \,\forall m_1.
           \end{cases}
\end{align*}
\label{Pg_encod_binary}
By the covering lemma, if $R_A' > I(X;U_1) + \delta(\epsilon'')$ then
$$\lim_{n \rightarrow \infty } Pr( \exists m_1 , \; (X^n,U_1^n(m_1)) \in T_{\epsilon''}^n(X,U_1)) = 1,$$
where $\delta(\epsilon'') \rightarrow 0$ as $\epsilon'' \rightarrow 0$.
Rate of the overall encoding is $R_A < R_A' + \delta(\epsilon'')$
for large enough $n$ such that
$Pr( \forall m_1 , \; (X^n,U_1^n(m_1)) \not\in T_{\epsilon''}^n(X,U_1))
< \delta(\epsilon'')/\log|\cX|$.
Thus, any rate $R_A > I(X;U_1) + 2\delta(\epsilon'')$ is sufficient.

Encoding at node B is similar to that of the encoding at node A.

\noindent \textbf{Encoding at relay}:

\revd{
If $R_C> I(W; U_1,U_2)$, but $R_C < \max\{I(X;U_1), I(Y;U_2)\}$, the
the relay uses the encoding as given in case 1 below. If
$R_C > \max\{I(X;U_1), I(Y;U_2)\}$, then the relay uses the encoding
as given in case 2.
}

\revd{\noindent \underline{\textbf{Case 1: $\max\{I(X;U_1), I(Y;U_2)\} > R_C > I(W; U_1,U_2)$}}}

The relay receives either an index $m_1$ or a $x^n$ sequence from node A. Similarly, from node B the relay receives
$m_2$ or a $y^n$ sequence. 
If $m_1$ and $m_2$ are received, and
$(u_1^n(m_1), u_2^n(m_2),w^n(m_3)) \in  T_{\epsilon}^n(U_1,U_2,W) $ for
some $m_3$, then any such $m_3$ is broadcasted by the relay.
In any other case, the relay broadcasts both the received sequences.
So the encoding at the relay is given by \footnote{\revd{This encoding
can be represented by a prefix-free code using standard techniques, as
outlined in the previous footnote.}}
\begin{align*}
 \phi_C & = \begin{cases}
               m_3  \quad (u_1^n(m_1),u_2^n(m_2), w^n(m_3)) \in T_{\epsilon}^n(U_1,U_2,W) \\
               (\phi_A(x^n), \phi_B(y^n)) \quad \mbox{ otherwise.}
           \end{cases}
\end{align*}
Let $E_{n,\epsilon'}$ be the event $(U_1^n(m_1),U_2^n(m_2)) \in T_{U_1U_2, \epsilon'}^n$ at the relay.
Then from the Markov lemma, we have $\lim\limits_{n \rightarrow \infty } Pr(E_{n,\epsilon'}) = 1$.
By the covering lemma, if $R_C' > I(W;U_1,U_2) + \delta(\epsilon)$ then
$$\lim\limits_{n \rightarrow \infty } Pr( \exists m_3 , \; (U_1^n(m_1), U_2^n(m_2),W^n(m_3))\in T_{\epsilon}^n(U_1,U_2,W) \;| \; E_{n,\epsilon'}) = 1,$$
where $\delta(\epsilon) \rightarrow 0$ as $\epsilon \rightarrow 0$.
Rate of the overall encoding is $R_C < R_C' + 2\delta(\epsilon)$
for large enough $n$ such that
$$Pr(E_{n,\epsilon'}\cap \forall m_3, (U_1^n(m_1), U_2^n(m_2),W^n(m_3))\notin T_{\epsilon}^n(U_1,U_2,W)) < \delta(\epsilon)/\log (\cU_1 \cdot \cU_2),$$
and $Pr\left(  E_{n,\epsilon'}^{c}) < \delta(\epsilon)/\log (a \cdot b \right)$, 
where $a = \max\{|\cX|, |\cU_1| \}$ and $b = \max\{|\cY|, |\cU_2|\}$.
Thus, any rate $R_C > I(W;U_1,U_2) + 3\delta(\epsilon)$ is sufficient.

\label{Pg_Scheme2}

\revd{
\noindent \underline{ \textbf{Case 2: $R_C > \max\{I(X;U_1), I(Y;U_2)\}$}}}

Let us consider the case where the relay
receives $m_1$ and $m_2$ from node A and B respectively,
such that $(u_1^n(m_1),u_2^n(m_2)) \in T_{U_1U_2, \epsilon'}^n$.
Then the relay broadcasts the XOR
of the binary representations of $m_1$ and $m_2$ 
(after padding zeros to the shorter sequence).
In any other case, as in scheme 1, the relay broadcasts both the 
received sequences. So the encoding at the relay is given by\footnote{\revd{This encoding
can be represented by a prefix-free code using standard techniques, as
outlined in the footnote in page~\pageref{Pg_encod_binary}.}}
\begin{align*}
 \phi_C & = \begin{cases}
               m_1 \oplus m_2  \quad (u_1^n(m_1),u_2^n(m_2)) \in T_{\epsilon'}^n(U_1,U_2) \\
               (\phi_A(x^n), \phi_B(y^n)) \quad \mbox{ otherwise.}
           \end{cases}
\end{align*}
By using the Markov lemma as before,
rate of the overall encoding is $R_C < \max\{R_A, R_B \} + 2\delta(\epsilon')$
for large enough $n$ such that
$Pr((U_1^n(m_1),U_2^n(m_2)) \notin T_{U_1U_2, \epsilon'}^n) < \delta(\epsilon')/\log (|\cU_1|.|\cU_2|)$.
Thus any rate $R_C > \max\{R_A, R_B \} + 2\delta(\epsilon')$ is sufficient.

\noindent \textbf{Decoding at node A}:

\revd{ If the relay follows the encoding scheme given in case 1, then node A performs the decoding procedure given in case 1 below. Othersiwe, it follows
the decoding operation given in case 2.
}

\revd{
\noindent \underline{ \textbf{Case 1:}}}

Node A receives either $m_3, m_2$ or $y^n$.
We show that node A  computes $f(x_i,y_i)$ with zero-error
 $\forall i$ for which $(x_i,y_i) \in S_{XY}$.
 Let us consider a pair $(x^n,y^n)$ such that
$(x_i,y_i) \in S_{XY}$ for some $i$.

\noindent \underline{{\it Subcase 1}: Node A receives $m_3$}

In this case, node A and node B had chosen $m_1$ and $m_2$ 
such that $(x^n, u_1^n(m_1)) \in T_{\epsilon''}^n(X,U_1) $  
and $(y^n, u_2^n(m_2)) \in T_{\epsilon''}^n(Y,U_2) $ respectively, 
and at the relay $(u_1^n(m_1),u_2^n(m_2), w^n(m_3)) \in T_{\epsilon}^n(U_1,U_2,W)$.


As a robustly typical sequence can not have a zero-probability
component (see \eqref{Eq_Typical}),
the sequence $u_1^n(m_1)$ chosen by node A satisfies $p(u_{1i}|x_i)>0 \; \forall i$,
since $(x^n, u_1^n(m_1)) \in T_{\epsilon''}^n(X,U_1)$.
Similarly, the sequence $u_2^n(m_2)$ chosen by node B satisfies $p(u_{2i}|y_i)>0 \; \forall i$,
and the sequence $w^n(m_3)$ chosen by the relay satisfies $p(w_i|u_{1i},u_{2i})>0\; \forall i$.
Hence $p(u_{1i}|x_i)p(u_{2i}|y_i)p(w_i|u_{1i},u_{2i})>0$.
Thus by Lemma~\ref{Sche2_lemma3}, if $p(x_i,y_i)>0$ then node A can 
compute  $f(x_i,y_i)$ from $x_i, u_{1i}$ and $w_i$ with zero-error.

\noindent \underline{{\it Subcase 2}: Node A receives $m_2$}

In this case node B had $m_2$ such that $(y^n, u_2^n(m_2)) \in T_{\epsilon''}^n(Y,U_2)$.
This shows that $p(u_{2i}(m_2)|y_i)>0$ for all $i$. 
Thus by Lemma~\ref{Lem_zero_schem1}, node A can recover $f(x_i,y_i)$ with zero-error
for all $i$ such that $p(x_i,y_i)>0$.


\noindent \underline{{\it Subcase 3}: Node A receives $y^n$}

In this case node A can compute $f(x_i,y_i)$ for all $i$.

\revd{
\noindent \underline{ \textbf{Case 2:}}}

Node A receives either $m_1 \oplus m_2, m_2$ or $y^n$. Let us first consider the case 
where node A receives $m_1 \oplus m_2$.
Since node A has $m_1$, it can decode $m_2$ from $m_1 \oplus m_2$ by XORing the
received message with $m_1$. Zero-error function computation at node A from $m_2$ and $x^n$
follows from Lemma~\ref{Lem_zero_schem1}.
Decoding for all other cases is the same as decoding in scheme 1.


Node B follows the similar decoding procedure as that of node A. 
Now using time sharing random variable $Q$ gives the achievability of every triple $(R_A,R_B,R_C)$
in $\cR_{I}$ for some $p(q)p(w|u_1,u_2,q)p(u_1|x,q)p(u_2|y,q)$. 
This completes the proof of part~(\ref{Thm_inner_part1}).
\hfill{\rule{2.1mm}{2.1mm}}

{\em Proof of part~(\ref{Thm_inner_part2}):}
\revd{For the function computation problem in Example~\ref{Ex_Zero_Pent},
we show that $\exists \; (U_1',U_2', W')$ s.t. $I(W';U_1',U_2') = \log 2$, and $ \max \{ I(X;U_1), I(Y;U_2)\} > \log 2$
for any choice of $(U_1,U_2)$. 
}
In this example, graphs $G_{X|Y}^f$ and $G_{Y|X}^f$ are pentagon
graphs.The complementary graph entropy of a pentagon graph with uniform distribution is shown to be
$\frac{1}{2} \log 5 $ \cite{Rose_2003}.
Since the graph entropy is greater than or equal to the complementary graph entropy,
we get $ \max \{ I(X;U_1), I(Y;U_2)\} > \frac{1}{2} \log 5 $
for any choice of $(U_1,U_2)$. 
Let us consider a scheme for the choice of $U_1 = \{X\}$ and $U_2 =\{Y\}$. Then $R_A = R_B = \log 5$. For this choice of $(U_1,U_2)$, the graph $\faux$
is same as the graph $\frooks$ which is shown in Fig.~\ref{Rook_graph}.
Let us choose 
$$
W = 
\begin{cases}
   \{(u_1,u_2)| u_1 = u_2 \} \mbox{ if } U_1 =U_2 \\
   
   \{(u_1,u_2)| u_1 \neq u_2 \} \mbox{ if } U_1 \neq U_2.
  \end{cases}
$$
Then $W$ is a binary random variable with uniform distribution
and satisfies all the conditions in Theorem~\ref{Thm_Zero_Inner1}.
\revd{
Here, since $W$ is a function of $(U_1,U_2)$, we get $I(W;U_1,U_2) = H(W) = \log 2$, and 
we have $ \max \{ I(X;U_1), I(Y;U_2)\} > \log 2 $.
Then we have the desired result.
}

\revd{
For the function computation
problem in Example~\ref{Aux_example}, we show that  $\exists \; (U_1',U_2')$ s.t. $ \max \{ I(X;U_1'), I(Y;U_2')\} = \frac{2}{3}$, and $I(W;U_1,U_2) > \frac{2}{3}$ for any $(U_1,U_2,W)$.
}
To prove this, we  consider the same of choices of $U_1$ and $U_2$
given in Section~\ref{Sec_zero_results}
and we use the following claim.
\begin{claim}
\label{Claim_Zero2}
 The only conditional distribution $p_{U_1|X}$ achieving
 $R_A = \frac{2}{3} $ for the function computation problem in
 Example~\ref{Aux_example}, is  $p_{U_1|X}(a|2) = p_{U_1|X}(b|2) = \frac{1}{2}$.
\end{claim}
\begin{IEEEproof}
 To prove the above claim, we need to show that $I(X;U_1)$ is strictly 
 convex in $p_{U_1|X}$. Let us take  $p_{U_1|X}(a|2) = p$, for $0 <p<1$.
 Then $I(X;U_1)$ is a function of $p$ which can be written as 
 $$I(X;U_1) = f(p) = -\frac{1}{3}(1+p) \log \frac{1}{3}(1+p) -\frac{1}{3}(2-p) \log \frac{1}{3}(2-p) + \frac{1}{3} p \log p + \frac{1}{3}(1-p) \log (1-p).$$
 Next we show that $f''(p) >0, \; \mbox{ for } 0<p<1$.
 \begin{align*}
 f'(p) = & \frac{1}{3} \log \frac{1}{3}(1+p) - \frac{1}{3} \log \frac{1}{3}(2-p) + \frac{1}{3} \log p - \frac{1}{3} \log (1+p). \\
 f''(p) = & \frac{1}{1+p} + \frac{1}{2-p} + \frac{1}{3}\frac{1}{p}+ \frac{1}{3}\frac{1}{1-p}.
 \end{align*}
Then we have $f''(p) > 0, \quad \mbox{ for } 0<p<1$. This proves the claim.
\end{IEEEproof}

For Example~\ref{Aux_example},
the confusability graphs $G_{X|Y}^f$ and $G_{Y|X}^f$ are the same and
it is shown in Fig.~\ref{Confus_Graph}.
For  uniform distribution on its vertices, the graph entropy of the graph shown in Fig.~\ref{Confus_Graph}, is computed as $\frac{2}{3}$ 
in Example 1 in \cite{Orlitsky_2001}. So we have $H_{G_{X|Y}^f}(X) = H_{G_{Y|X}^f}(Y) = \frac{2}{3}$. 
Then we get $(\frac{2}{3},\frac{2}{3},\frac{2}{3}) \in \cR_{I2}$.
Claim~\ref{Claim_Zero2} shows that we have to choose  $p_{U_1|X}(a|2) = p_{U_1|X}(b|2) = p_{U_2|Y}(c|2) = p_{U_2|Y}(d|2) = \frac{1}{2}$ to achieve the rates $R_A=R_B= \frac{2}{3} $.
For this choice of $(U_1,U_2)$, let us compute the joint distribution of $(U_1,U_2)$.
Note that $(U_1,U_2) = (a,c)$ has non zero joint probability with $(X,Y)$ when either $(X,Y) = (1,2)$
or $(X,Y) = (2,1)$. By marginalizing over $(X,Y)$, we get $p_{U_1,U_2}(a,c) = \frac{1}{6} $.
Similarly, we get 
$ p_{U_1,U_2}(b,d) = \frac{1}{6},  p_{U_1,U_2}(a,d) = p_{U_1,U_2}(b,c) = \frac{1}{3}$.

As we have seen before the  graph $\faux$ is a ``square'' graph which is 
shown in Fig.~\ref{Aux_graph}. The minimum $R_C$ achievable by Scheme 1 
in this case is $H_{G_{U_1U_2}^f}(U_1,U_2)$. 
For this graph $\faux$, the only two maximal independent sets are $  \{(a,c),(b,d)\}$ and  $\{(a,d),(b,c)\} $.
Let $W$ be a random variable distributed over $\{\{(a,c),(b,d)\},\{(a,d),(b,c)\} \} $.
Since each node of the graph $\faux$ 
is contained in only one of the maximal independent set,
we have $w$ as a function of $(u_1,u_2)$. Then
$R_C = I(W;U_1,U_2) = H(W) = H(\frac{1}{3}) \approx 0.91$. 
This shows that for the above choice of $(U_1,U_2)$, the minimum $R_C$
achievable using scheme 1 is $ H(\frac{1}{3})$.
Since $ H(\frac{1}{3}) > \frac{2}{3}$, we get $(\frac{2}{3},\frac{2}{3},\frac{2}{3}) \notin \cR_{I1}$. 
This completes the proof of part~(\ref{Thm_inner_part2}).
\hfill{\rule{2.1mm}{2.1mm}}

\subsection{Proof of Theorem~\ref{Relay_function}}
\label{Sec_Relay_knows}
\revd{Since $p(x,y)>0$ $ \forall \, (x,y) \in \cX \times \cY$,
$p(x^n, y^n) > 0$ for any $x^n \in \cX^n$ and $y^n \in \cY^n$.
Let $f^n(x^n,y^n)$ denote $(f(x_1,y_1),\cdots, f(x_n,y_n))$.
If the relay node cannot compute the function with zero-error, it implies that there exists $x^n, y^n, x'^n, y'^n$ 
such that $\phi_A(x^n) = \phi_A(x'^n), \phi_B(y^n) = \phi_B(y'^n)$ and $f^n(x^n, y^n) \neq f^n(x'^n, y'^n)$. 
We either have $f^n(x^n, y^n) \neq f^n(x^n, y'^n)$ or $f^n(x^n, y'^n) \neq f^n(x'^n, y'^n)$.
W.l.o.g., let us assume that $f^n(x^n, y^n) \neq f^n(x^n, y'^n)$. 
For both pairs $(x^n, y^n)$ and $(x^n, y'^n)$, node A receives
$\phi_C(\phi_A(x^n), \phi_B(y^n))$ from the relay. 
Then node A cannot compute the function since the relay's message and $X^n$
are the same for these pairs, but the values of the function are different. So we get a contradiction.
This proves the result. 
\hfill{\rule{2.1mm}{2.1mm}}
\label{Pg_relay_knows}
}

\section{$\epsilon$-error computation: Proofs of Theorems~\ref{Thm_Epsilon_Inner1}-~\ref{Thm_Rooks} }
\label{Sec_asym_bounds}
{\em Proof of Lemma~(\ref{Lem_Asym_Outer_Bound}),  part~(\ref{Lem_Asym_Outer1})}:
Let us consider the cut between node A and the super-node consisting of B and C. 
Then it is the function computation problem with side information considered
in \cite{Orlitsky_2001} where the decoder with side information $Y$ 
wants to compute a function $f(X,Y)$. They showed that the optimal 
$\epsilon$-error rate for this problem is $H_{G_{X|Y}^{f}}(X|Y)$.
This implies that $R_A \geq H_{G_{X|Y}^{f}}(X|Y)$. Similarly, $R_B \geq H_{G_{Y|X}^{f}}(Y|X)$.
The lower bound for $R_C$ follows from the cut set bound by considering the 
cut $(\{C\},\{A,B\})$ and assuming that the relay knows $(X,Y)$.

{\em Proof of part~(\ref{Lem_Asym_Outer2})}:
Let us consider a scheme where nodes A and B 
encode $X^n$ and $Y^n$ to messages $m_1$ and $m_2$
by the scheme 
given by Orlitsky and Roche in \cite{Orlitsky_2001}, and the relay 
broadcasts both these messages. From the result of \cite{Orlitsky_2001}, 
$R_A = H_{G_{X|Y}^{f}}(X|Y)$,$R_B=H_{G_{Y|X}^{f}}(Y|X)$, and
$R_C=H_{G_{X|Y}^{f}}(X|Y)+ H_{G_{Y|X}^{f}}(Y|X)$ are 
achievable using this scheme.
Now let us consider another scheme where $X$ and $Y$ are communicated to the relay from A and B.
Then the relay first computes $f(X,Y)$ and then uses Slepian-Wolf binning
to compress it at a rate $R_C = \max\{H(Z|X), H(Z|Y)\}$. 
Then nodes A and B can compute $f(X,Y)$
with negligible probability of error.
The rates $R_A = H(X), R_B=H(Y)$, and
$R_C=\max\{H(Z|X), H(Z|Y)\}$ are achievable for this scheme.
\hfill{\rule{2.1mm}{2.1mm}}

\subsection{Proof of Theorem~\ref{Thm_Epsilon_Inner1}}
\label{Asym_Inner}


{\em Proof of part~(\ref{Thm_asym_inner_part1})}:
The scheme used to prove the achievability of $\eR_{I1}^{\epsilon }$ is
similar to that of $\cR_{I}$ in Theorem~\ref{Thm_Zero_Inner1}.
Nodes A and B follow Berger-Tung coding scheme \cite{Berger_1977}.
(We refer the reader to Theorem~12.1 in \cite{Elgamal_Kim}.)
At node A, like in scheme 1 in Theorem~\ref{Thm_Zero_Inner1},
a codebook  $\{ U_1^n(m_1) |m_1 \in\{ 1,\cdots,{2^{nR_A'}}\}\}$
is used. The codebook is randomly binned into 
$2^{nR_A}$ bins. If a $u_1^n(m_1)$ is found which
is jointly typical with $x^n$, then its bin index $b_1$ is sent. If
no such $u_1^n$ is found in the codebook, then a randomly chosen bin 
index is sent.  Node B encodes in a similar way.

The relay can correctly recover $(m_1,m_2)$ from $b_1$ 
and $b_2$ with high probability if $ R_A  > I(X;U_1|U_2)$, 
$R_B  > I(Y;U_2|U_1)$ and $ R_A + R_B  > I(X,Y; U_1, U_2)$.
Let the reconstructed messages be $(\hat{m}_1,\hat{m}_2)$.
The relay follows Wyner-Ziv coding scheme  where a codebook
$\{ W^n(m_3) |m_3 \in \{ 1, \cdots, {2^{nR_C'}}\}\}$
is randomly binned into $2^{nR_C}$ bins. If the relay finds a $w^n(m_3)$ which
is jointly typical with $(u_1^n(\hat{m}_1),u_2^n(\hat{m}_2))$,
then it broadcasts the bin index of $w^n(m_3)$. Otherwise a randomly chosen
bin index is broadcasted.
Node A can decode $m_3$ correctly with high probability if
\begin{align*}
R_C \stackrel{(a)}{=} & I(W;U_1,U_2) - I(W;X,U_1) + \epsilon\\
    = & H(W|X,U_1) - H(W|U_1,U_2) + \epsilon\\
    \stackrel{(b)}{=}  & H(W|X,U_1) - H(W|U_1,U_2,X) + \epsilon\\
    = & I(W;U_2|X,U_1)  + \epsilon.
\end{align*}
Here in $(a)$, we have taken the size of the bin as $2^{n(I(W;X,U_1)+\epsilon')}$, and
$(b)$ follows from the Markov chain $W-U_1U_2-X$.
Similarly node B can decode $m_3$ with high probability if 
$R_C \geq I(W;U_1|Y,U_2)$. 

Let the reconstructed messages at nodes A and B be $\hat{m}_3^A $ and $\hat{m}_3^B$
respectively. 
Then $w^n(\hat{m}_3^A)$ will be jointly typical with $(x^n, u_1^n(m_1), y^n, u_2^n(m_2))$
with high probability. For such a $w^n(\hat{m}_3^A)$, for all $i$ 
such that $p(x_i,y_i)>0$,
we get  $p(u_{1i}|x_i)p(u_{2i}|y_i)p(w_i|u_{1i},u_{2i})>0$  using robust typicality.
Thus by Lemma~\ref{Sche2_lemma3}, node A can 
compute  $f(x_i,y_i)$ from $x_i, u_{1i}$ and $w_i$.
Node B computes the function in a similar way.

Now let us consider the encoding schemes used to obtain the rate region $ \eR_{I2}^{\epsilon}$.
Node A encodes $X^n$ to an index $m_1$ using the scheme given 
by Orlitsky and Roche in \cite{Orlitsky_2001}.
Using the same scheme, node B encodes $Y^n$ to an index $m_2$ 
with rate $R_B$.
Once the relay receives both the messages, it broadcasts the XOR of 
the binary representation of $m_1$ and $m_2$ (after appending zeros to the shorter sequence).
Nodes A recovers message $m_2$ from $(m_1,m_1 \oplus m_2)$.
Then node A follows the decoding operation given in \cite{Orlitsky_2001}
to compute the function. Similar decoding operation is performed at node B.
By the result of \cite{Orlitsky_2001}, 
$R_A = H_{G_{X|Y}^{f}}(X|Y)$,$R_B=H_{G_{Y|X}^{f}}(Y|X)$, and
$R_C=\max \{H_{G_{X|Y}^{f}}(X|Y), H_{G_{Y|X}^{f}}(Y|X)\}$ are 
achievable using this scheme.
\hfill{\rule{2.1mm}{2.1mm}}

{\em Proof of part~(\ref{Thm_asym_inner_part2})}:
Let us consider computing $X\cdot Y$ (AND function)
for DSBS($p$) $(X,Y)$. Here both the confusability graphs $G_{X|Y}^{f}$
and $G_{Y|X}^{f}$ are complete. This implies
$H_{G_{X|Y}^{f}}(X|Y) = H(X|Y)$ and
$H_{G_{Y|X}^{f}}(Y|X)=H(Y|X)$.
Since $H(X|Y) = H(Y|X) = H(p)$, we get
$\eR_{I2}^{\epsilon } = \{(R_A,R_B,R_C) : R_A, R_B, R_C  \geq H(p) \}$.
Now let us consider the achievable scheme of  
$\eR_{I1}^{\epsilon}$ in Theorem~\ref{Thm_Epsilon_Inner1} for this example.
Since both the confusability graphs are complete, 
the only choice for $U_1$ and $U_2$ are $U_1 = \{X\}$ and $U_2 = \{Y\}$.
For this choice of $U_1$ and $U_2$, the relay can recover $X$ and $Y$ 
by Berger-Tung coding scheme.
Then the relay can compute the function $Z = f(X,Y)$.
For a given $Z=z$, let us consider the set of all $(x,y)$, $A_z=\{(x,y): f(x,y)=z, \mbox{ and } (x,y)\in S_{XY} \}$.
Let us choose $W= A_{Z}$. Then we get $R_C = \max\{H(Z|X),H(Z|Y)\} = \frac{1}{2}H(p)$, which is 
the minimum possible $R_C$ by Lemma~\ref{Lem_Asym_Outer_Bound}.
So we get  $\eR_{I1}^{\epsilon}$ as 
\begin{align*}
 & \{(R_A,R_B,R_C) : R_A \geq  H(p), R_B \geq H(p), R_A + R_B \geq 1+H(p), R_C  \geq \frac{1}{2}H(p) \}. 
\end{align*}
Then we have  $(H(p),H(p),H(p)) \in \revd{\eR_{I2}^{\epsilon} \setminus \eR_{I1}^{\epsilon}}$
and  $(1,H(p),\frac{1}{2} H(p)) \in \revd{\eR_{I1}^{\epsilon} \setminus \eR_{I2}^{\epsilon}} $.
\hfill{\rule{2.1mm}{2.1mm}}

\subsection{Proof of Theorem~\ref{Thm_Rooks}}
\label{Sec_Asym_Rooks}
We use the following lemma to prove Theorem~\ref{Thm_Rooks}.
For $f_1,f_2$ of $(X,Y)$, let the random variables $Z_1$ and $Z_2$ 
denote $f_1(X,Y)$ and $f_2(X,Y)$ respectively.
\begin{lemma}
\label{Lem_Rooks}
 If $E(\frookA) \subseteq E(\frookB)$, then $H(Z_1|Z_2,X) = 0$ and $H(Z_1|Z_2,Y) = 0$. 
\end{lemma}
\begin{IEEEproof}
 We prove that if $E(\frookA) \subseteq E(\frookB)$, then 
 $H(Z_1|Z_2,X) = 0$. The other case follows similarly.  
 For a given $X=x$ and $Z_2=z_2$, let us consider the set 
 of all $y$, $A_{xz_2} = \{y': f_2(x,y') = z_2 \mbox{ and } (x,y')\in S_{XY} \} $.
 Then by the definition of $\frookB$, $f_2(x,y') = f_2(x,y'')$ $\forall y',y'' \in A_{xz_2}$.
 Further, since $E(\frookA) \subseteq E(\frookB)$, $f_1(x,y') = f_1(x,y'')$.
 Let us denote this unique value by $z_1 := f_1(x,y')$.
 Then we have $Pr\{Z_1=z_1|X=x, Z_2=z_2\} = 1$ and $H(Z_1|Z_2,X) = 0$.
\end{IEEEproof}

{\em Proof of part~(\ref{Thm_rooks_part1}}):
Lemma~\ref{Lem_Rooks} shows that if  $E(\frookA) \subseteq E(\frookB)$, then
$Z_1$ is a function of $(Z_2,X)$ as well as a function of $(Z_2,Y)$. This implies that 
if node A can recover $Z^n_2$ from $M_C$ and $X^n$ with some probability of error,
then it can compute $Z_1^n$ with at most the same probability of error. Similar arguments hold 
for computing $Z_1^n$ at node B. This shows that 
$\RNzerunA \supseteq \RNzerunB$ and
$\RNepsA \supseteq  \RNepsB$.
\hfill{\rule{2.1mm}{2.1mm}}

Part~(\ref{Thm_rooks_part2}) follows from part~(\ref{Thm_rooks_part1}).

\section{Conclusion}
\label{Sec_conclusion}
In this work, we studied the function computation problem in a bidirectional relay
network (Fig.~\ref{Relay_Network}). Function computation problem has been 
addressed from an information theoretic point of view for unidirectional networks before,
e.g. \cite{Korner_1979,Han_1987,Orlitsky_2001,Shayevitz_2014}.
To the best of our knowledge, this is the first work which addressed
the function computation problem for a bidirectional network from an 
information theoretic point of view. We considered our function computation problem
on this network for correlated sources under zero-error and $\epsilon$-error criteria
and proposed single-letter inner and outer bounds for 
achievable rates.
We studied the function computation problem in a
broadcast network (Fig.~\ref{Broadcast_network}), where we showed
that the optimal broadcast rate is the same
under zero-error and $\epsilon$-error criteria.


\appendices

\rev{
\section{Source coding under zero-error vs. $\epsilon$-error}
We now mention a result which connects zero-error with $\epsilon$-error. We believe this result is folklore. We provide it here for completeness. The result in the following lemma can be extended/applied to our source coding network.
	\begin{lemma}
		\label{Clm_rate_eqv} Let us consider a source
coding problem with side information where the encoder knows $X$ and the 
decoder has the side
information $Y$ and wants to recover $X$. If there is a zero-error prefix
free code of rate $R$, then for any $\delta >0$, the rate $R+\delta$ is achievable under $\epsilon$-error.
	\end{lemma}
	\begin{proof}
		Consider a zero-error prefix free code of length $n$ and rate $R$. Let  $M_A$ denote the 
encoded message.
		Since the average length of any prefix free encoding is lower bounded by the entropy of the source, we get $nR \geq H(M_A)$.
		Now let us consider a block encoding of $N$ messages
$M_A$ under $\epsilon$-error.
		For any $\delta >0$, there exists an $N$ such that by random binning of $M_A^N$ symbols at a rate $H(M_A)+\delta$, the decoder can reconstruct $M_A$ with arbitrarily small probability of error.  
		Since the source vectors $X^n$ can be reconstructed with zero-error
from $M_A$,  the decoder can decode $X$  with arbitrarily small probability of
error. This proves the lemma.  \end{proof}
}

\end{document}